\documentclass[sigconf, nonacm]{acmart}

\newcommand\vldbyear{2026}
\newcommand\vldbworkshop{AIDB}
\newcommand\vldbauthors{\authors}
\newcommand\vldbtitle{\shorttitle}
\newcommand\vldbavailabilityurl{https://github.com/BauplanLabs/DBPLANBENCH}
\newcommand\vldbpagestyle{plain}

\usepackage{amsmath}
\usepackage{mathtools}
\usepackage{amsthm}

\usepackage{algorithm}
\usepackage{algorithmic}
\usepackage{float}
\floatstyle{ruled}
\restylefloat{algorithm}

\usepackage[capitalize,noabbrev]{cleveref}

\newcommand{\name}{{\color{mediumpurple}\textsc{DBPlanBench}}}

\usepackage{listings}

\definecolor{mediumpurple}{rgb}{0.58, 0.44, 0.86}
\definecolor{MEDIUMPURPLE}{rgb}{0.58, 0.44, 0.86} 

\usepackage{tikz}
\usetikzlibrary{backgrounds}
\usetikzlibrary{arrows.meta,positioning}

\tikzset{
  joinop/.style={draw, rounded corners, align=center, font=\scriptsize, inner sep=2.6pt, line width=0.7pt},
  leaf/.style={draw, rounded corners, align=center, font=\scriptsize, inner sep=2.4pt, line width=0.6pt},
  build/.style={-Latex, very thick, black},
  probe/.style={-Latex, thick, dashed, gray!65},
  changed/.style={fill=gray!15, draw=black, line width=0.9pt},
}

\lstdefinelanguage{json}{
    basicstyle=\ttfamily\small,
    showstringspaces=false,
    breaklines=true,
    literate=
     *{:}{{{\color{red}:}}}{1}
      {,}{{{\color{red},}}}{1}
      {\{}{{{\color{blue}\{}}}{1}
      {\}}{{{\color{blue}\}}}}{1}
      {[}{{{\color{blue}[}}}{1}
      {]}{{{\color{blue}]}}}{1}
}

\theoremstyle{plain}

\theoremstyle{definition}

\theoremstyle{remark}

\usepackage{graphicx}
\usepackage[most]{tcolorbox}
\tcbuselibrary{skins,breakable,listings}
\usepackage{multirow}
\usepackage{siunitx}
\tcbuselibrary{listings,breakable,skins}

\newtcblisting{promptblock}[2][]{
  breakable,
  enhanced,
  colback=black!2,
  colframe=black!35,
  arc=2pt,
  boxrule=0.5pt,
  left=6pt,right=6pt,top=4pt,bottom=4pt,
  title={#2},
  fonttitle=\bfseries,
  listing only,
  listing options={
    language={},
    basicstyle=\ttfamily\footnotesize,
    columns=fullflexible,
    keepspaces=true,
    breaklines=true,
    showstringspaces=false,
    literate={→}{{$\rightarrow$}}1 {⨝}{{$\bowtie$}}1
  },
  #1
}

\newtcblisting{jsonblock}[2][]{
  breakable,
  enhanced,
  colback=black!2,
  colframe=black!35,
  arc=2pt,
  boxrule=0.5pt,
  left=6pt,right=6pt,top=4pt,bottom=4pt,
  title={#2},
  fonttitle=\bfseries,
  listing only,
  listing options={
    language=json,
    basicstyle=\ttfamily\footnotesize,
    columns=fullflexible,
    keepspaces=true,
    breaklines=true,
    breakatwhitespace=true,
    showstringspaces=false,
  },
  #1
}

\lstdefinelanguage{SQL}{
  morekeywords={
    SELECT,FROM,WHERE,GROUP,BY,HAVING,ORDER,AS,JOIN,INNER,LEFT,RIGHT,FULL,OUTER,ON,
    WITH,UNION,ALL,DISTINCT,CASE,WHEN,THEN,ELSE,END,AND,OR,NOT,NULL,IS,IN,EXISTS,
    SUM,AVG,MIN,MAX,COUNT,RANK,DENSE_RANK,ROW_NUMBER,OVER,PARTITION,ROWS,RANGE,
    LAG,LEAD,COALESCE,NULLIF,CAST,DATE,EXTRACT,INTERVAL,LIKE,LIMIT,OFFSET
  },
  sensitive=false,
  morecomment=[l]{--},
  morestring=[b]',
}

\lstdefinestyle{sqlpretty}{
  language=SQL,
  basicstyle=\ttfamily\scriptsize,
  columns=fullflexible,
  keepspaces=true,
  breaklines=true,
  breakatwhitespace=true,
  showstringspaces=false,
  tabsize=2,
  upquote=true,
  numbers=left,
  numberstyle=\ttfamily\tiny,
  numbersep=6pt,
  xleftmargin=1.2em,
  keywordstyle=\bfseries\color{black!75},
  stringstyle=\color{black!60},
  commentstyle=\itshape\color{black!45},
}

\definecolor{tpchaccent}{HTML}{CC887A}
\definecolor{tpcdsaccent}{HTML}{7A88CC}

\colorlet{tpchmix}{mediumpurple!40!tpchaccent}
\colorlet{tpcdsmix}{mediumpurple!40!tpcdsaccent}

\colorlet{tpchbg}{tpchmix!10!white}
\colorlet{tpchframe}{tpchmix!60!black}

\colorlet{tpcdsbg}{tpcdsmix!10!white}
\colorlet{tpcdsframe}{tpcdsmix!60!black}

\newtcblisting{tpchquery}[2][]{
  breakable,
  enhanced,
  colback=tpchbg,
  colframe=tpchframe,
  arc=2pt,
  boxrule=0.5pt,
  left=6pt,right=6pt,top=4pt,bottom=4pt,
  title={#2},
  fonttitle=\bfseries\small,
  listing only,
  listing options={
    style=sqlpretty
  },
  #1
}

\newtcblisting{tpcdsquery}[2][]{
  breakable,
  enhanced,
  colback=tpcdsbg,
  colframe=tpcdsframe,
  arc=2pt,
  boxrule=0.5pt,
  left=6pt,right=6pt,top=4pt,bottom=4pt,
  title={#2},
  fonttitle=\bfseries\small,
  listing only,
  listing options={
    style=sqlpretty
  },
  #1
}

\usepackage{subcaption}
\usepackage{xcolor}
\usepackage[normalem]{ulem}

\newlength{\algcommentwidth}
\newlength{\algcodewidth}
\newlength{\algcommentprobe}
\newlength{\algcommentcol}

\renewcommand{\algorithmiccomment}[1]{
  \setlength{\algcommentcol}{0.45\linewidth}
  \settowidth{\algcommentprobe}{\footnotesize$\triangleright$~#1\ }
  \ifdim\algcommentprobe>\algcommentcol
    \unskip\hfill\penalty-10000
    \hspace*{3.5em}
    \begin{minipage}[t]{\dimexpr\linewidth-4em\relax}
      \footnotesize\raggedright
      \setlength{\parindent}{0pt}
      \leftskip=1em \hangindent=-1em \hangafter=1
      \noindent\llap{$\triangleright$~}#1\par
    \end{minipage}
    \vspace{0.6ex}
  \else
    \quad\hbox to \dimexpr\algcommentcol-\algcommentprobe\relax{}
    {\footnotesize$\triangleright$~#1}
  \fi
}

\providecommand{\citet}[1]{\citeauthor{#1}~\cite{#1}}

\newtcolorbox{schemabox}[2][]{
  lower separated=false,
  colback=white,
  colframe=mediumpurple,
  fonttitle=\bfseries\small,
  colbacktitle=mediumpurple,
  coltitle=white,
  enhanced,
  attach boxed title to top left={yshift=-0.07in,xshift=0.1in},
  boxed title style={boxrule=0pt,colframe=white},
  title=#2,#1}

\newtcolorbox{tracebox}[2][]{breakable,enhanced,colback=black!2,colframe=black!35,arc=2pt,boxrule=0.5pt,left=6pt,right=6pt,top=4pt,bottom=4pt,title={#2},fonttitle=\bfseries,#1}

\newcommand{\elision}{\\[2pt]\null\hfill{\color{black!45}[\,\dots\,]}\hfill\null\\[2pt]}

\begin{document}
\title{Test-Time Optimization of Physical Query Plans with LLMs}

\author{Mehmet Hamza Erol}
\affiliation{
  \institution{Stanford University}
  \city{Stanford}
  \state{CA}
  \country{USA}
}
\email{mhamza@stanford.edu}

\author{Xiangpeng Hao}
\affiliation{
  \institution{University of Wisconsin-Madison}
  \city{Madison}
  \state{WI}
  \country{USA}
}
\email{xiangpeng.hao@wisc.edu}

\author{Federico Bianchi}
\affiliation{
  \institution{TogetherAI}
  \city{San Francisco}
  \state{CA}
  \country{USA}
}
\email{federico@together.ai}

\author{Ciro Greco}
\affiliation{
  \institution{Bauplan}
  \city{New York}
  \state{NY}
  \country{USA}
}
\email{ciro.greco@bauplanlabs.com}

\author{Jacopo Tagliabue}
\affiliation{
  \institution{Bauplan}
  \city{New York}
  \state{NY}
  \country{USA}
}
\email{jacopo.tagliabue@bauplanlabs.com}

\author{James Zou}
\affiliation{
  \institution{TogetherAI / Stanford University}
  \city{San Francisco / Stanford}
  \state{CA}
  \country{USA}
}
\email{jamesz@stanford.edu}

\begin{abstract}
Traditional query optimization relies on cost-based optimizers that estimate execution cost (e.g., runtime, memory, and I/O) using predefined heuristics and statistical models. Improving these requires substantial engineering effort, yet they often cannot exploit semantic correlations in queries and schemas that could enable better physical plans. Large language models (LLMs), however, can reason about column semantics, value distributions, and broader domain context that classical statistics miss. We introduce \name, a harness for the DataFusion engine that exposes physical plans through a compact serialized representation and applies LLM-proposed edits as JSON patches. On this harness, we instantiate a test-time optimization workflow where an LLM examines physical query plans, proposes localized edits based on semantic reasoning, and an evolutionary search refines the candidates across iterations. We target OLAP queries, where heavy, repeated execution turns even small efficiency gains into substantial cumulative savings. We specifically focus our evaluation on join reordering and join-side selection, where cardinality-estimation errors compound multiplicatively. Median speedups reach $1.10$-$1.12\times$ on TPC-H and $1.05$-$1.07\times$ on TPC-DS, with some achieving up to $4.78\times$. We also demonstrate that optimizations discovered at small scale factors transfer effectively to larger ones, supporting a low-cost small-to-large workflow.
\end{abstract}

\maketitle

\pagestyle{\vldbpagestyle}
\begingroup\small\noindent\raggedright\textbf{VLDB Workshop Reference Format:}\\
\vldbauthors. \vldbtitle. VLDB \vldbyear\ Workshop: \vldbworkshop.\\ 
\endgroup
\begingroup
\renewcommand\thefootnote{}\footnote{\noindent
This work is licensed under the Creative Commons BY-NC-ND 4.0 International License. Visit \url{https://creativecommons.org/licenses/by-nc-nd/4.0/} to view a copy of this license. For any use beyond those covered by this license, obtain permission by emailing \href{mailto:info@vldb.org}{info@vldb.org}. Copyright is held by the owner/author(s). Publication rights licensed to the VLDB Endowment. \\
\raggedright Proceedings of the VLDB Endowment. 
ISSN 2150-8097. \\
}\addtocounter{footnote}{-1}\endgroup

\vspace{-3mm}

\ifdefempty{\vldbavailabilityurl}{}{
\vspace{.3cm}
\begingroup\small\noindent\raggedright\textbf{VLDB Workshop Artifact Availability:}\\
The source code, data, and/or other artifacts have been made available at \url{\vldbavailabilityurl}.
\endgroup
}

\section{Introduction}

\begin{figure}[t]
\vspace{4mm}
  \centering
  \includegraphics[width=1\columnwidth]{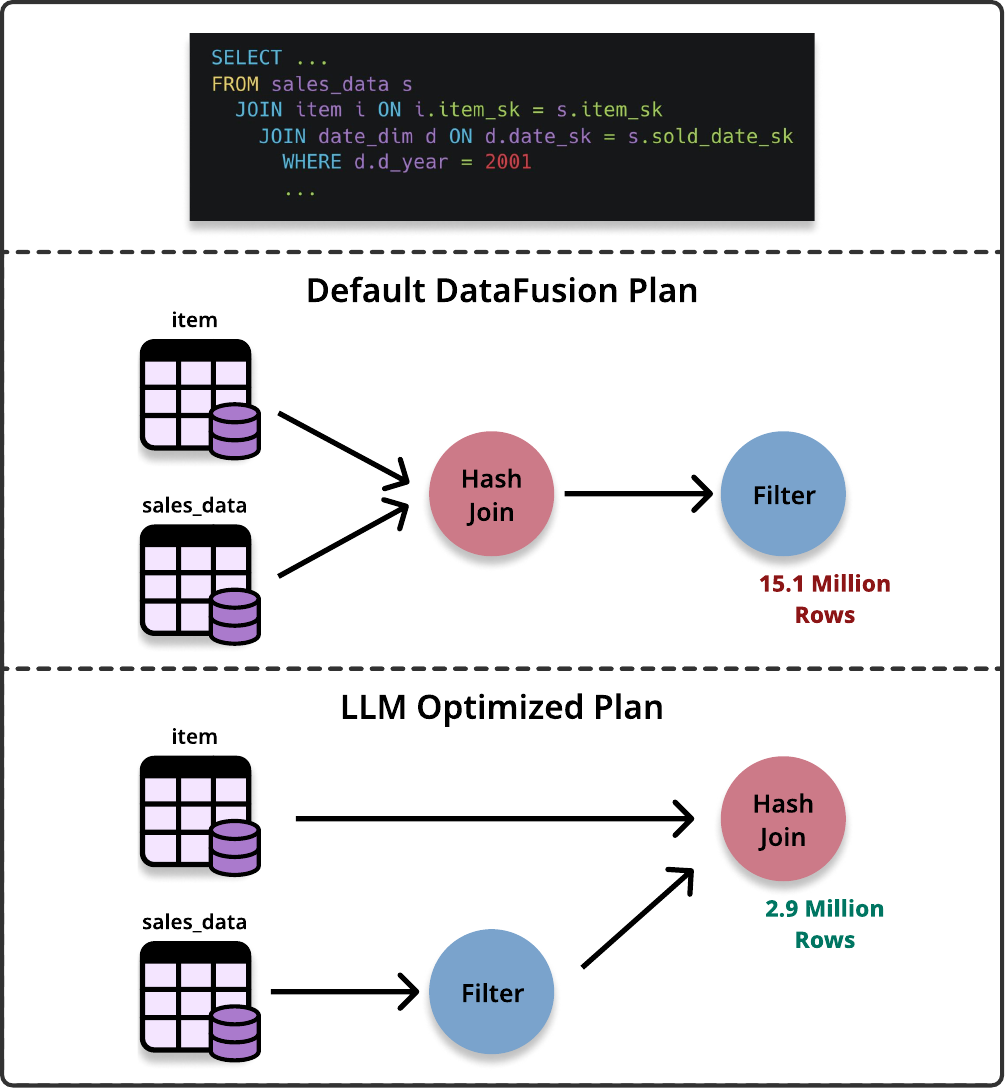}
  \caption{Semantics-aware Join Optimization. Exploiting the high selectivity of the `date' predicate, the plan is reordered to prune data early, reducing the cardinality of the subsequent \texttt{JOIN}. We explore this in Section~\ref{case-study}.}
  \label{fig:case_study_viz}
\end{figure}

\begin{figure*}[t]
  \centering
  \includegraphics[width=0.95\textwidth]{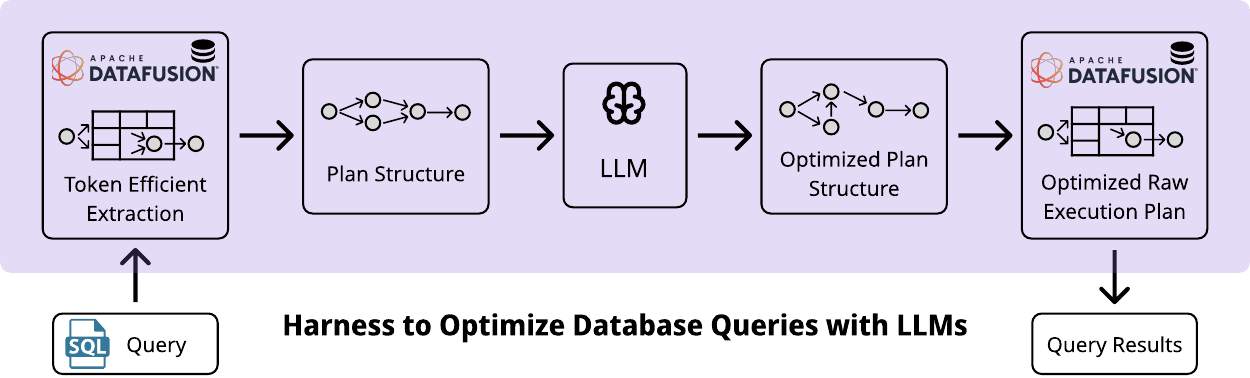}
  \caption{\name{} provides accessible APIs to extract a physical plan that an LLM can edit. This plan can be then executed in DataFusion as if it were a native DataFusion plan.}
  \label{fig:pipeline_overview}
\end{figure*}

Large Language Models' (LLMs) ability to generate code to solve verifiable tasks has been the key to unlocking a new paradigm in automated optimization, with preliminary, yet encouraging results in model inference \cite{cheng2025barbariansgateaiupending}, GPU kernels \cite{ouyang2025kernelbenchllmswriteefficient}, cloud scheduling policies \cite{tagliabue2025aidistributedsystemsdesign}, and mathematical reasoning \cite{nagda2025reinforcedgenerationcombinatorialstructures}. OLAP (Online Analytical Processing) workloads form a substantial segment of the cloud database market \cite{cloud_dw_market}, with a verifiable task at their core, \textit{query planning}: a SQL engine needs to turn a declarative user query into a series of physical operations (a physical plan) that are both semantically correct (i.e., the results match user intent) and efficient (i.e., the query is processed as fast as possible). OLAP queries are heavily reused~\cite{van2024tpc,wu2024stage,10825377}, so even small improvements in planning would translate into meaningful efficiency gains at cloud scale~\cite{cloud_dw_market}.

Given their success in similar settings, it is therefore natural to ask: \textit{can LLMs discover better physical plans when compared to algorithms crafted by human experts}? For example, a reasonable hypothesis is that LLMs could use their ``semantic'' knowledge to optimize the order of a \texttt{JOIN} operation, making a query faster~\cite{lohman2014query}. As illustrated in Figure~\ref{fig:case_study_viz}, an LLM can leverage semantic context to infer that the `date' predicate is highly selective, allowing it to reduce cardinality earlier than a standard optimizer.

Using an off-the-shelf LLM (GPT-5), we instantiate a test-time optimization framework with an evolutionary sampling strategy to optimize an existing physical execution plan. We show that physical plans discovered by the model can reach 4.78$\times$ speedups, opening up a wide variety of optimization and cost savings for queries.\footnote{Considering that the API calls needed for substantial improvements in our test cases usually cost only a few cents, real-world workloads with repeated and overlapping queries can achieve a positive ROI from automated optimization. Appendix~\ref{sec:cost_benefit} quantifies this, showing time-based crossover within typical OLAP-template reuse counts.}

To this end, we develop \name, a harness designed to evaluate OLAP planning in production environments (Figure~\ref{fig:pipeline_overview}). We integrate \name~on Apache DataFusion~\cite{lamb2024apache}, a widely adopted, industrial-grade query engine that has emerged as a backbone of modern analytical systems, powering production deployments across time-series databases, observability platforms, cloud SQL services, and serverless lakehouses~\cite{tagliabueduckhunt,safronovr2sql,dixinfluxdb3,datafusionusers}. DataFusion exemplifies the move toward composable data systems~\cite{pedreira2023composable} and serves as a rigorous baseline, as its query optimizer incorporates decades of database research representing the current standard for non-learned optimization. Thus, we unpack the DataFusion database architecture to expose its internal API to end users. We then intercept query execution to retrieve the physical plan, serialize it into a compact representation, and provide it to an LLM to discover semantically guided optimizations. We demonstrate the usefulness of our test-time optimization pipeline using evolutionary sampling on two benchmarks derived from TPC-H~\cite{tpc-h} and TPC-DS~\cite{tpc-ds}. We show that LLMs can make query execution as much as 4.78$\times$ faster and we further demonstrate that the optimized plans found on smaller DBs generalize to larger DBs effectively and maintain the speedups. The harness and all the code used in the experiments will be released as open-source. We summarize our contributions as follows:

\begin{itemize}
\item We introduce \name{}, a test harness that exposes database physical plans to LLMs via a novel, token-efficient serialization format and a patch-based editing interface, enabling direct structural refinement of execution plans.
\item We run evolutionary sampling as a test-time optimization strategy on this harness, showing that LLMs can optimize query plans, outperforming the industrial-grade DataFusion optimizer.
\item We show that LLMs can perform semantic reasoning about cardinalities, using domain knowledge to identify non-obvious structural optimizations, such as join reordering and build-side swapping, that rule-based heuristics miss.
\item We validate a small-to-large workflow for cross-scale test-time optimization, showing that optimizations discovered cheaply at small scale factors can be reliably transferred to larger scales via rule-based procedures, preserving correctness and retaining meaningful speedups. \end{itemize}

We further support our setup and findings with cross-model and real-world robustness experiments, a patch-vs-full-generation ablation, and an evolutionary-convergence analysis (Section~\ref{sec:analysis}), complemented by timing-variance and external-verification checks (Appendix~\ref{apdx:verification}).
\section{\name}

\label{sec:dbplanbench}
A core contribution of our work is designing \name{}, a harness for studying plan optimization in a practical and effective manner (Figure~\ref{fig:pipeline_overview} gives an overview of the pipeline). A primary obstacle to LLM-driven optimization is that query plans in DataFusion are volatile, in-memory objects internal to the engine, lacking a standardized external interface. We address this by introducing a dedicated serialization layer that performs a deep traversal of the engine's physical operator graph. This layer maps heterogeneous engine-specific objects, such as join strategies and data partitioning schemes, into a unified JSON schema. We also implement Python APIs to execute physical JSON plans directly.

However, a direct mapping is insufficient; a naive serialization results in an ``information explosion'' that overwhelms the LLM's context window. Consequently, our serialization layer filters out redundant information while preserving the semantic properties for an LLM to make meaningful optimizations.
Appendix~\ref{sec:exec_plan_details} provides full technical details via an example, including the raw plan serialization, the compressed representation used by the LLM, and a visualization of the join graph before and after patching.

\subsection{Plan Execution and Representation}

Modern query engines execute SQL by compiling it into a \emph{plan}: a directed operator graph that specifies how data is read, transformed, and combined to produce the final result. This compilation proceeds in two major optimization stages.
First, the engine translates SQL into a \emph{logical plan} and applies \emph{logical optimizations}, such as expression simplification, predicate pushdown, projection pruning, and operator reordering. These transformations are semantic-preserving rewrites that improve efficiency without committing to specific execution strategies.
Second, the optimized logical plan is converted into a \emph{physical plan}, where concrete execution choices are made. Physical optimization selects join algorithms, scan methods, data partitioning strategies, and degrees of parallelism. The resulting physical plan is the final, executable specification that governs query execution.

\name~operates \emph{after} this traditional optimization pipeline (Figure~\ref{fig:pipeline_overview}). Rather than replacing the query optimizer, it allows exposing the plan itself, thereby allowing an LLM to refine an already-optimized physical plan.

To enable this, \name~first serializes the in-memory physical plan into JSON. However, this first raw serialization is extremely verbose: it contains full file paths, detailed type encodings, repeated per-partition metadata, and other execution-specific details that are both irrelevant to optimization and expensive in LLM context. Feeding such data directly to an LLM is inefficient and quickly exhausts the context window.

We therefore introduce a \emph{streamlined physical plan representation} that preserves all optimization-relevant information while aggressively removing redundancy.\footnote{For instance, without this compression, a median TPC-DS query's prompt would grow roughly ten-fold to over 2 million characters (Figure~\ref{fig:comp_vs_length}, Table~\ref{tab:prompt_component_breakdown}), thereby exceeding standard context windows and making optimization of complex workloads infeasible.} This representation deduplicates file-level statistics, compresses type encodings, and omits execution-irrelevant fields such as file locations and per-partition replicas. The result is typically about 10$\times$ smaller than the original serialized plan, making it practical to provide as LLM input. Here is an example of reduction, where information coming from different columns is aggregated in a single statistic.

\begin{schemabox}{Statistics Representation}
{\scriptsize
\textbf{Full Stats (24 lines/col $\times$ 14 cols = 336 lines)}
\begin{verbatim}
"columnStats": [{
  "distinctCount": {
    "precisionInfo": "ABSENT", "val": {}},
  "maxValue": {"val": {"int64Value": "6"}},
  "minValue": {"val": {"int64Value": "1"}},
  "nullCount": {"val": {"uint64Value": "0"}},
  "sumValue": {
    "precisionInfo": "ABSENT", "val": {}}
}, ... ]  // repeated for each column
\end{verbatim}
}
\tcblower
{\scriptsize
\textbf{Succinct Stats (4 lines total)}
\begin{verbatim}
"statistics": {
  "n_rows": 6,
  "total_bytes": 1243
}
\end{verbatim}
}
\end{schemabox}

\subsection{Plan Edits Representation}

Even with the streamlined plan representation, generating entirely new physical plans is often token-intensive (see Section~\ref{sec:comp_vs_speedup}, paragraph on prompt lengths), which increases the likelihood of structural errors and parsing failures. In practice, however, critical optimizations, such as JOIN swaps, are typically localized and do not require a full rewrite.

Thus, rather than asking the LLM to emit an entirely new optimized plan, we instead have it produce a set of \emph{JSON Patch} operations. These patches describe localized edits to the input plan and are orders of magnitude smaller than the full plan representation. \name~applies the patches to the streamlined plan to obtain an optimized version. An example of a JSON patch for a JOIN swap application is given next:

\begin{schemabox}{JSON Patch: Swap Join Sides}
{\scriptsize
\textbf{Before}
\begin{verbatim}
'37': {'hashJoin': {'left': 38, 'right': 80,
'on': [{'left': 'item_sk', 'right': 'i_item_sk'}]}}
\end{verbatim}
}

\tcbline
{\scriptsize
\textbf{Patch}
\begin{verbatim}
[{'op': 'replace', 'path': '/left', 'value': 80},
 {'op': 'replace', 'path': '/right', 'value': 38},
 {'op': 'replace', 'path': '/on/0/left',
  'value': 'i_item_sk'},
 {'op': 'replace', 'path': '/on/0/right',
  'value': 'item_sk'}]
\end{verbatim}
}

\tcbline
{\scriptsize
\textbf{After}
\begin{verbatim}
'37': {'hashJoin': {'left': 80, 'right': 38,
'on': [{'left': 'i_item_sk', 'right': 'item_sk'}]}}
\end{verbatim}
}
\end{schemabox}

Finally, the optimized streamlined plan is re-expanded into the full physical plan format expected by the DataFusion engine and executed normally.

\section{Optimizing Query Plans with LLMs}
\begin{figure*}[h]
  \centering
  \includegraphics[width=\textwidth]{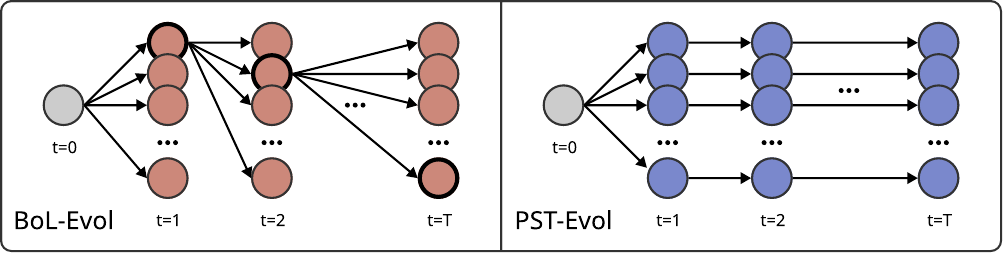}
  \caption{Visualization of the two evolutionary algorithms, BoL-Evol (left) and PST-Evol (right), detailed in Algorithm~\ref{alg:evolution}. Gray nodes: original plan; colored nodes: modified plans; arrows: LLM-generated patches; bold-edged nodes (BoL-Evol only): best plan at each step.}
  \label{fig:algorithm_viz}
\end{figure*}

We optimize physical plans with LLMs relying on their ability to reason and apply semantics-aware transformations. We use GPT-5 as our optimization model, ask it to (i) analyze a given physical plan node-by-node, (ii) estimate intermediate cardinalities using semantic knowledge of schemas and predicates, and (iii) generate concrete plan transformations that reduce execution cost while preserving correctness.

\subsection{LLM-Guided Plan Transformations}

\paragraph{\textbf{Semantic Cardinality Estimation.}} We instruct the LLM to use semantic knowledge about column names, table contexts, and real-world data distributions to estimate cardinalities. For example, when estimating the selectivity of a filter on an \texttt{order\_date} column, the LLM can reason about temporal patterns in business data rather than relying solely on uniform distribution assumptions. The full prompt is provided in the Appendix.

The LLM performs cardinality estimation bottom-up through the plan tree:
\begin{itemize}
\item \textbf{ParquetScan nodes}: Use the exact row count from table statistics
\item \textbf{Filter nodes}: Analyze filter predicates semantically to estimate selectivity based on column semantics (temporal, ID, status, demographic columns)
\item \textbf{Join nodes}: Estimate join cardinalities based on relationship types (foreign keys, many-to-many) and typical join ratios observed in real-world data
\end{itemize}

\textbf{Plan Transformations.} Based on cardinality estimates, we ask the LLM to propose two primary types of optimizations:

\begin{enumerate}
\item \textbf{Join-Side Selection}: Hash joins in DataFusion build a hash table from the left input and probe with the right input. Performance is optimal when the smaller relation is on the left. The LLM identifies cases where $|left| > |right|$ and generates patches to swap the inputs.

\item \textbf{Join Reordering}: In multi-join queries, performing lower-cardinality joins first reduces the size of intermediate results that must be processed by subsequent joins. The LLM identifies beneficial reorderings, transforming plans from $(A \bowtie B) \bowtie C$ to $(A \bowtie C) \bowtie B$ when $|A \bowtie C| < |A \bowtie B|$.
\end{enumerate}

We focus on these two transformations because cardinality estimation errors compound multiplicatively across joins, noted as the dominant source of plan suboptimality in classical analyses~\cite{lohman2014query}. The harness supports other physical-plan optimizations through the same patch interface, which we discuss in Section~\ref{sec:limitations}.

When generating transformations, the LLM must maintain several invariants to ensure not changing the original execution's operational intent: 1) Preserve all nodes (no deletions or disconnections) 2) Maintain valid DAG topology (no cycles) 3) Update projection indexes when swapping join inputs to reflect the new schema ordering 4) Update join conditions and metadata consistently with structural changes

\subsection{DataFusion Plans and Execution}

DataFusion generates physical plans represented as directed acyclic graphs (DAGs) where nodes represent physical operators (ParquetScan, Filter, HashJoin, Aggregate, etc.) and edges represent data flow.

We serialize these plans to JSON format where each node is assigned a unique ID and contains fields specific to its operator type. For example, a HashJoin node contains \texttt{left} and \texttt{right} fields referencing its input node IDs, an \texttt{on} field specifying join conditions, and a \texttt{projection} field determining which columns to output.

Our system intercepts plans after DataFusion's physical optimization phase but before execution, allowing us to modify the physical plan while still benefiting from DataFusion's rule-based logical optimizations.

\subsection{Execution and Evaluation Protocol}
We run all SQL executions in our pipeline under a common sandboxed evaluation protocol to keep measurements comparable across runs. We use Modal\footnote{\url{https://modal.com/}} to launch a containerized environment from a prebuilt image that pins our DataFusion build and required Python / Rust dependencies and loads the Parquet datasets into an in-memory filesystem to minimize I/O variability, which is crucial for obtaining stable runtime estimates. Each sandbox is provisioned with fixed resources (4 CPUs and 4GB of RAM) to keep measurements comparable across runs.

Given a candidate plan serialized to JSON, we launch a fresh sandbox, reconstruct the plan inside DataFusion, and measure end-to-end execution time. We repeat this across \(R=50\) independent sandboxes and use the minimum observed latency as our timing estimate, aligning with robust benchmarking practice under intermittent runtime noise where sandbox-level effects predominantly introduce transient slowdowns rather than speedups.\footnote{\citet{chen2016robust} argue that because environmental noise only adds delay, benchmark distributions are right-skewed, making \texttt{min} the most robust estimator of noise-free performance. Appendix~\ref{apdx:sensitivity} validates this under our setting: \texttt{min} yields lower cross-group variability (CoV across (R{=}50) independent timing groups) than \texttt{mean} or \texttt{median}.}  Finally, we terminate each sandbox after the run to avoid cross-run state (e.g., caching) influencing measurements.

\subsection{Algorithmic Framework}
For test-time optimization of physical plans with LLMs, we employ evolutionary sampling. The main idea is to restructure sampling into several steps, where the generated transformations at a step evolve the ones from the prior step. We continue this process until observing convergence.

We implement two evolutionary frameworks: (1) Parallel Single Threads Evolution (\texttt{PST-Evol}) and (2) Best of Last Evolution (\texttt{BoL-Evol}), as shown in Figure~\ref{fig:algorithm_viz} and in Algorithm~\ref{alg:evolution}. In \texttt{PST-Evol}, we generate \(K\) improved candidate plans from the original query plan, then evolve each candidate independently for a total of \(T\) steps, drawing one sample from the model at each step. In \texttt{BoL-Evol}, we maintain a single ``current best'' plan: starting from the original plan, we sample \(K\) improved candidates, keep the best, and repeat this process for a total of \(T\) steps.

In addition, we include a \texttt{Best-of} sampling baseline, where we draw $T \times K$ candidate plans from the base LLM and compare the best-performing one against the plans obtained via evolutionary optimization.
\section{Experiments}

\sisetup{
  detect-weight = true,
  detect-inline-weight = math,
  table-number-alignment = center
}

\begin{table*}[t]
\centering
\small
\setlength{\tabcolsep}{8pt}
\begin{tabular}{
  l c
  S[table-format=1.2] S[table-format=1.2] S[table-format=1.2] S[table-format=1.2] S[table-format=1.2]
  S[table-format=2.2] S[table-format=2.2] S[table-format=2.2] S[table-format=2.2] S[table-format=2.2]
}
\toprule
\multirow{2.5}{*}{Dataset} & \multirow{2.5}{*}{Method}
& \multicolumn{5}{c}{\(\text{speedup}_q(T)\) over \(q\in Q\)}
& \multicolumn{5}{c}{\(\text{fast}_{Q,s}(T)\) for thresholds \(> s\) (\%)} \\
\cmidrule(lr){3-7} \cmidrule(lr){8-12}
& & {P25} & {P50} & {P75} & {P90} & {Max}
& {\(>1.00\)} & {\(>1.05\)} & {\(>1.10\)} & {\(>1.25\)} & {\(>1.40\)} \\
\midrule

\multirow{3}{*}{TPC-H}
& \texttt{Best-of}
& {\bfseries 1.05} & 1.10 & 1.16 & 1.23 & 1.56
& {\bfseries 91.67} & {\bfseries 74.17} & 51.67 & 7.50 & {\bfseries 2.50} \\
& \texttt{BoL-Evol}
& 1.04 & {\bfseries 1.12} & {\bfseries 1.19} & {\bfseries 1.27} & {\bfseries 1.67}
& 90.83 & 73.33 & {\bfseries 55.83} & {\bfseries 11.67} & {\bfseries 2.50} \\
& \texttt{PST-Evol}
& 1.04 & 1.10 & 1.17 & 1.25 & 1.62
& 90.00 & 71.67 & 50.83 & 10.00 & 1.67 \\
\midrule

\multirow{3}{*}{TPC-DS}
& \texttt{Best-of}
& {\bfseries 1.03} & {\bfseries 1.07} & 1.10 & 1.13 & 3.02
& {\bfseries 87.50} & {\bfseries 64.17} & 20.83 & 1.67 & 0.83 \\
& \texttt{BoL-Evol}
& 1.02 & 1.05 & 1.09 & 1.12 & 4.37
& 89.17 & 52.50 & 15.83 & 1.67 & 0.83 \\
& \texttt{PST-Evol}
& {\bfseries 1.03} & 1.06 & {\bfseries 1.11} & {\bfseries 1.18} & {\bfseries 4.78}
& 85.83 & 60.83 & {\bfseries 29.17} & {\bfseries 4.17} & {\bfseries 1.67} \\
\bottomrule
\end{tabular}

\caption{Summary Statistics of Per-query Speedups at the Final Round for Query Complexities 5--8. We use \(T{=}4\) and \(K{=}5\). \texttt{BoL-Evol}/\texttt{PST-Evol} evolve for \(T\) steps with \(K\) samples per step; \texttt{Best-of} draws \(T\times K\) independent samples and reports the best.}
\label{tab:speedup_summary}
\vspace{-5mm}
\end{table*}

\begin{figure*}[t]
  \centering
  \includegraphics[width=\linewidth]{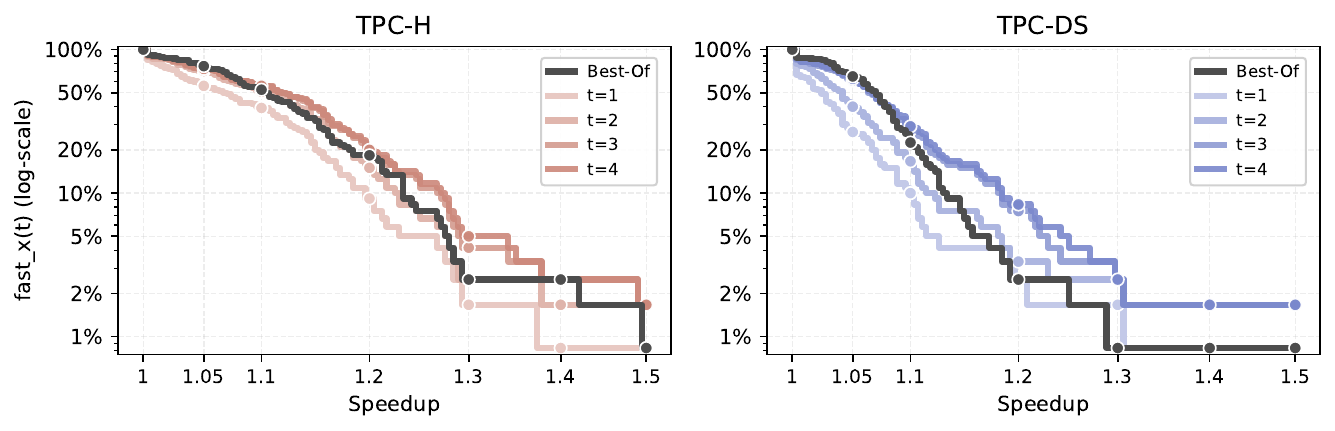}
  \caption{CCDF Plots Representing the Change in the Distribution Throughout the Evolution. \texttt{BoL-Evol} results are plotted for TPC-H and \texttt{PST-Evol} results are plotted for TPC-DS. The y-axis ($\text{fast}_x(t)$) is on a log-scale. We observe diminishing returns when the step is larger. Moreover, evolutionary algorithms help push the frontier in achieving higher speedups in contrast to base sampling.}
  \label{fig:ccdf_sep}
\end{figure*}

\paragraph{\textbf{Datasets.}} We evaluate our approach on workloads derived from two industry-standard decision support benchmarks: TPC-H~\cite{tpc-h} and TPC-DS~\cite{tpc-ds}.\footnote{Data generation was performed using the DuckDB TPC extensions: \url{https://duckdb.org/docs/current/core_extensions/overview}.} TPC-H consists of a schema with eight tables modeling the management of product sales and distribution. TPC-DS utilizes a complex snowflake schema comprising 24 tables, including 7 fact tables (modeling sales and returns across store, catalog, and web channels) and 17 dimension tables. Both benchmarks support variable dataset sizes via a scaling factor; we utilize a scale factor of 3 for all experiments.

\paragraph{\textbf{Query Suite Construction for \name.}} We use GPT-5 to generate SQL queries at controlled complexity levels. To ensure validity, we rigorously filter for queries that are syntactically correct, executable, and return non-empty and deterministic results.\footnote{While access to large-scale production workloads is often limited, LLM-generated query suites produce workloads with complexity and cost profiles representative of production systems~\cite{lao2025sqlbarber}. Furthermore, constructing custom suites is standard practice for isolating specific system behaviors that fixed benchmarks may not expose~\cite{schmidt2025sqlstorm}.} The prompt contains information about the schema of the different tables, and we define query complexity by the number of joins and aggregations required, highlighted with details in Appendix~\ref{sec:complexity_desc}.

For each dataset, we generate 30 queries at each complexity level in ${5,6,7,8}$, yielding 120 queries per dataset and 240 queries total. Queries in this range typically include more than three complex joins, nested subqueries, window functions, CTEs, and aggregation constructs such as GROUP BY and HAVING. Finally, we keep only queries that successfully execute, return a non-empty result under the default DataFusion optimizer, and run no shorter than $0.15$ seconds and no longer than $15$ seconds under our execution time estimation protocol.

\paragraph{\textbf{Methods and Baselines.}} Throughout our experiments, we sample $K=5$ per step for both evolutionary algorithms (\texttt{PST-Evol} and \texttt{BoL-Evol}), and run for $T=4$ steps. We compare our evolutionary approach against the default DataFusion optimizer, which employs rule-based transformations. This represents a strong baseline as DataFusion's optimizer incorporates decades of database optimization research. Additionally, we experiment with \texttt{Best-of} sampling with the base LLM, and contrast with the performance of the algorithms.

\paragraph{\textbf{Evaluation Metrics.}} We treat a candidate plan as \emph{valid} iff the patched plan deserializes, executes successfully, and produces the same output as the baseline query. Output equivalence gives a strict and automatable correctness signal but presumes deterministic query outputs, which we enforce in our query suite filter. We evaluate valid plans using end-to-end execution time. For each query $q$, let $p_0$ denote the baseline DataFusion plan with latency $\tau(p_0)$. After $t$ evolutionary steps, let $p_t^*$ be the best \emph{valid} plan observed up to round $t$ (including $p_0$). Our primary metric is per-query execution-time speedup,
\[
\textit{speedup}_{q}(t) \;=\; \frac{\tau(p_0)}{\tau(p_t^*)}.
\]
Following prior work~\cite{ouyang2025kernelbenchllmswriteefficient}, we also report the fraction of queries achieving more than the target speedup $s$,
\[
\text{fast}_{Q,s}(t) \;=\; \frac{1}{|Q|}\sum_{q \in Q}\mathbf{1}\!\left[\text{speedup}_{q}(t)> s\right],
\]
where $Q$ is a query set and $\mathbf{1}[\cdot]$ is the indicator function. In particular, $\text{fast}_{Q,1}(t)$ corresponds to the percentage of queries in $Q$ for which our approach finds a plan strictly faster than the baseline by step $t$ (i.e., $\tau(p_t^*)<\tau(p_0)$). Moreover, since invalid or non-improving candidates are filtered and the baseline plan is retained as the fallback (Algorithm~\ref{alg:evolution}), every reported per-query speedup is $\geq 1\times$ by construction, and the framework cannot create slower plans.

\section{Results}

\paragraph{\textbf{LLMs can optimize physical plans}}
We present an overview of the speedup statistics in Table~\ref{tab:speedup_summary}. For the evolutionary algorithms, we report the best optimization discovered up to step~$T$, while for \texttt{Best-of}, we consider the best plan among $T \times K$ samples. We observe that the breadth of \texttt{Best-of} helps achieve stronger speedups in the lower percentiles of queries, whereas the depth of evolutionary algorithms pushes the upper percentiles further. We see evolutionary algorithms complement each other across datasets, with \texttt{BoL-Evol} performing particularly well on TPC-H and \texttt{PST-Evol} on TPC-DS. Overall, we observe a substantial level of optimizability: median speedups are around  $1.1\times$-$1.12\times$ for TPC-H and around $1.05\times$-$1.07\times$ for TPC-DS. Moreover, we observe significant per-query improvements, reaching up to $1.67\times$ on TPC-H and $4.78\times$ on TPC-DS. The fast metric further shows that most speedups remain below $1.25$ for both datasets, and that for roughly $10\%$ of queries no improvement is found. Speedups above $1.4$ are rare. Taken together, these results indicate that GPT-5 is effective at optimizing physical execution plans and can yield meaningful performance improvements. Speedups show no strong dependence on query characteristics. Across the features we examine, the pooled and TPC-DS trends are weakly negative at most, most clearly for window-function and CTE counts ($\rho\approx-0.3$ on TPC-DS), and no single characteristic reliably predicts the gain (Appendix~\ref{sec:speedup_vs_char}).

\paragraph{\textbf{LLMs use semantic knowledge to optimize plans}}\label{case-study}

We analyze the TPC-DS query with the maximum observed speedup ($4.78\times$, see Figure~\ref{fig:case_study_viz}). The query performs cross-channel sales analysis, joining sales data from three channels with \texttt{item}, \texttt{customer\_address}, and \texttt{date\_dim} tables, filtered by \texttt{d\_year = 2001}. Traditional optimizers produce a suboptimal plan here because, lacking precise statistics for the year filter, they rely on heuristics that prioritize joining the smaller \texttt{item} table (36K rows) before the larger \texttt{date\_dim} table (73K rows).

However, the LLM uses semantic reasoning to infer that a single year in a century-scale dimension is highly selective, making \texttt{date\_dim} the most effective filter to apply first. The optimization process identifies this opportunity and rewires the plan DAG using a 16-operation JSON patch. Appendix~\ref{sec:anatomy} walks through this optimization, including the model's reasoning trace (Appendix~\ref{sec:cot_attribution}), 
the serialized plan and the applied patch. The trace shows the model estimating the cardinalities bottom-up and inferring the year's selectivity from calendar knowledge, and updating the build-side and filtering order accordingly.

More specifically, the optimization applies two key transformations. First, join reordering moves the date join earlier in the pipeline, filtering 80\% of rows ($15.1M \rightarrow 2.9M$) before downstream joins. Second, join-side selection swaps inputs to build hash tables from dimension tables (36K + 94K rows) instead of larger fact tables. Figure~\ref{fig:tpcds-top-join-graph} shows the join graph before and after these two changes.

The optimized plan significantly reduces resource consumption. It lowers total build memory by $3.3\times$, peak per-task build memory by $1.6\times$, and the summed hash-join output rows by $1.9\times$ (per-step improvements detailed in Appendix~\ref{sec:cot_attribution}). The reduction in output rows is the most telling, as it is engine-independent and is the quantity classical cost models tie query cost to~\cite{lohman2014query, leis2015job}. These reductions directly contribute to the $4.78\times$ end-to-end speedup. This pattern is not unique to this case study, as we show it holds across the top-30 queries (Figure~\ref{fig:metric_ratios_vs_speedup}), illustrating that semantic analysis can unlock optimizations the engine misses.
\vspace{-1mm}
\paragraph{\textbf{Novel plans transfer to larger-scale databases}}
\label{sec:scale_transfer}

\begin{figure}[t!]
  \centering
  \includegraphics[width=0.9\columnwidth]{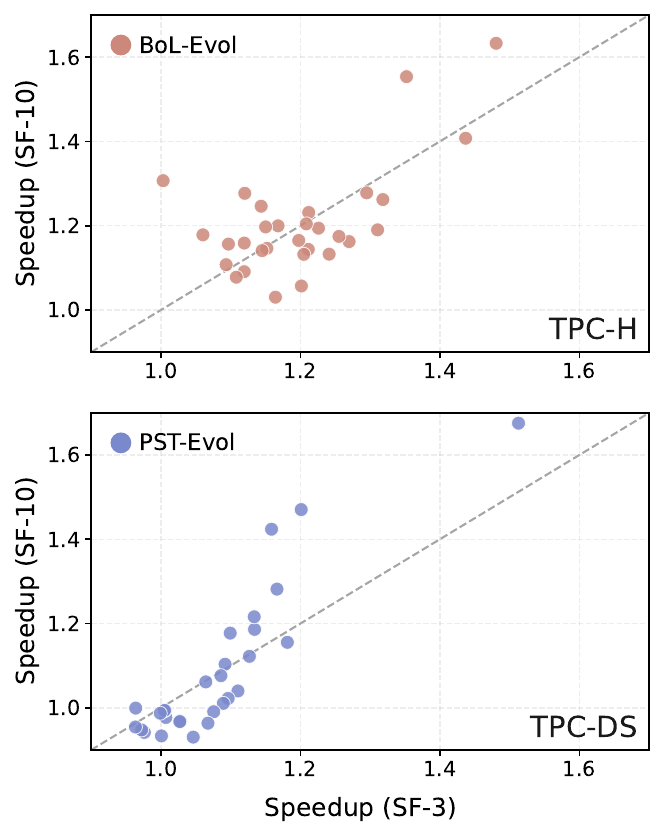}
  \caption{Plan-transfer generalization across scale factors.
  Scatter of speedup at large scale vs.\ small scale under the transfer method, evaluated at sandboxes with 16 GB memory. The diagonal indicates parity ($y=x$). One TPC-DS outlier (SF3 $=3.82$, SF10 $=3.83$) is omitted.}
  \label{fig:scale_transfer}
  \vspace{-2mm}
\end{figure}

Our main optimization results are obtained at scale factor 3 (SF3), which provides a prototypical setting where we can iterate quickly while still exercising non-trivial OLAP behavior. A natural question is whether the optimizations discovered at SF3 remain useful at larger scales. This question is practically important: in many real deployments, the cost of repeated exploration at full scale is prohibitive, so one would like to (i) discover candidate optimizations at a smaller scale and (ii) reliably apply them at a larger scale. This transfer protocol is a natural use case for test-time optimization, where the search runs once at a cheap proxy scale and its result is deployed at the target scale.

For each dataset, we study transfer for the top 25\% by SF3 speedup, taking the best plan found for each query. We then evaluate transfer to a larger scale (here SF10) by first obtaining the \emph{baseline} physical plan at the target scale and running a deterministic transfer procedure that produces a runnable \emph{optimized} physical plan structure for the target scale (see Appendix~\ref{app:transfer_details} for implementation details and limitations). We re-evaluate these runs using the same timing and correctness protocol as in the main experiments, with one exception: we increase the memory limit to 16 GB at both scales to accommodate larger-scale executions, which also ensures memory is matched across scales and keeps the SF3 and SF10 speedups comparable.

Our runs on both TPC-H and TPC-DS benchmarks successfully transfer all selected optimized plans to larger scale. Moreover, as Figure~\ref{fig:scale_transfer} shows, their SF3 and SF10 speedups are positively correlated, preserving the gains found at the smaller scale.  We present this as a feasibility study rather than a deployment guarantee, and scale is only one way the optimization and deployment settings can differ. Since the target-scale baseline is always available, a transferred plan can be validated against it before use, and serious deployments should confirm speedups and correctness at the target scale first. Overall, these results demonstrate that optimizing at a smaller scale can be a useful proxy and testbed for improving performance at larger scales.

\section{Analysis}
\label{sec:analysis}

\paragraph{\textbf{Convergence and Sample Efficiency.}}
\label{sec:convergence}

We analyze the convergence of the optimization by the algorithms in Figure~\ref{fig:ccdf_sep}. We observe in Figure~\ref{fig:ccdf_sep} that until step $T=3$, evolution visibly helps push the frontier of the discovered speedups, but the returns diminish at step $T=4$. Compared against \texttt{Best-of} at the same $T\times K$ samples, evolution pushes the upper percentiles further, while \texttt{Best-of} is stronger at the lower percentiles. Evolution is also more sample-efficient, reaching \texttt{Best-of}'s full-budget median with fewer samples on TPC-H (roughly $1.33\times$) and remaining comparable on TPC-DS. We examine these comparisons in more detail in Appendix~\ref{sec:best_of_n}.

\paragraph{\textbf{Cross-Model Generality.}}
To assess whether the improvements depend on the choice of language model, we rerun the full PST-Evol configuration ($T{=}4$, $K{=}5$) on the top-30 TPC-DS queries with the open-source Kimi K2.5~\cite{kimiteam2026k25}, holding the prompts, queries, and execution pipeline fixed. Table~\ref{tab:cross_llm} reports the per-query speedups. The two models produce comparable speedup distributions and validity rates (GPT-5: $71.8\%$, Kimi K2.5: $66.7\%$), with GPT-5 the stronger model overall, indicating that the improvements arise from the optimization paradigm rather than from a particular model.

\begin{table}[t]
\centering
\small
\begin{tabular*}{0.9\columnwidth}{@{\extracolsep{\fill}}lcccccc}
\toprule
Model & P25 & P50 & P75 & P90 & Max & $>\!1\times$ \\
\midrule
GPT-5 & 1.149 & 1.187 & 1.246 & 1.294 & 4.782 & 100.0\% \\
Kimi K2.5 & 1.037 & 1.177 & 1.214 & 1.329 & 4.159 & 76.7\% \\
\bottomrule
\end{tabular*}
\caption{Cross-model generality on the top-30 TPC-DS queries under PST-Evol ($T{=}4$, $K{=}5$), with per-query wall-clock speedup (Modal, $R{=}50$, min estimator). The speedup distributions are comparable, though GPT-5 outperforms Kimi K2.5, and both reach similar validity ($71.8\%$ and $66.7\%$).}
\label{tab:cross_llm}
\vspace{-1mm}
\end{table}

\paragraph{\textbf{Robustness to Real-World Queries.}}
To confirm the improvements are not an artifact of LLM-generated queries and to measure speedups on real-world workloads, we evaluate on canonical, expert-written queries, namely $15$ randomly sampled reference queries from each of TPC-H and TPC-DS, and one randomly sampled variant of each of $33$ Join-Order Benchmark (JOB) query structures over the real IMDB dataset~\cite{leis2015job}, again under PST-Evol ($T{=}4$, $K{=}5$). As shown in Table~\ref{tab:real_world}, we observe positive median speedups on all three benchmarks ($1.156$ on TPC-H, $1.085$ on TPC-DS, and $1.228$ on IMDB/JOB), with $60$--$73\%$ of queries improved and substantial tail gains, reaching $6.07\times$ on IMDB.\footnote{\label{fn:job16}For JOB structure 16, whose variants exceed the default $4$\,GB sandboxes, we separately run all four variants on $8$\,GB sandboxes; they reach a median speedup of $2.29\times$ and a max of $6.60\times$, supporting our robustness findings.}

\begin{table}[t]
\centering
\small
\begin{tabular*}{0.9\columnwidth}{@{\extracolsep{\fill}}lccccccc}
\toprule
Dataset & $n$ & P25 & P50 & P75 & P90 & Max & $>\!1\times$ \\
\midrule
TPC-H & 15 & 1.014 & 1.156 & 1.411 & 1.462 & 1.714 & 73.3\% \\
TPC-DS & 15 & 1.000 & 1.085 & 1.246 & 1.568 & 2.137 & 60.0\% \\
IMDB (JOB) & 32 & 1.000 & 1.228 & 1.392 & 1.642 & 6.070 & 68.8\% \\
\bottomrule
\end{tabular*}
\caption{Per-query wall-clock speedup (Modal, $R{=}50$, min estimator, SF-3) under PST-Evol ($T{=}4$, $K{=}5$) on canonical expert-written queries, showing that our speedups persist on real-world queries. For JOB we sample one random variant per query structure (except structure 16; see footnote~\ref{fn:job16})}
\label{tab:real_world}
\vspace{-5mm}
\end{table}

\paragraph{\textbf{Patch vs.\ Full Generation.}}
We ablate the central design decision of \name{} by having the model generate a complete plan from scratch, as in end-to-end approaches, instead of editing the existing plan with patches. Both conditions use GPT-5 at a single step ($T{=}1$, $K{=}5$) on the top-30 TPC-DS queries and differ only in the plan editing interface. For full generation, we adapt the patch prompt to instruct the model to emit a complete plan rather than a patch. As shown in Table~\ref{tab:full_vs_patch}, full generation attains comparable speedups, slightly higher at the upper percentiles, but at a lower validity rate ($54\%$ vs.\ $70\%$) and a substantially higher total token cost (median $23{,}218$ vs.\ $8{,}492$ output tokens). This gap comes not only from the far larger emitted plan (median $7{,}289$ vs.\ $146$ patch tokens) but also from more reasoning, as producing a complete plan from scratch is a heavier task than proposing a localized edit. We observe that patch-based editing is more reliable and cheaper, while full generation remains a promising direction for future work.

\begin{table}[t]
\centering
\footnotesize
\begin{tabular*}{0.9\columnwidth}{@{\extracolsep{\fill}}llccccc}
\toprule
Metric & Method & P25 & P50 & P75 & P90 & Max \\
\midrule
\multirow{2}{*}{Speedup} & Patch & 1.000 & 1.005 & 1.052 & 1.206 & 2.694 \\
 & Full-gen & 1.000 & 1.002 & 1.070 & 1.245 & 3.799 \\
\midrule
\multirow{2}{*}{\#tokens (out)} & Patch & 106 & 146 & 226 & 332 & 779 \\
 & Full-gen & 4{,}837 & 7{,}289 & 9{,}669 & 14{,}201 & 23{,}223 \\
\midrule
\multirow{2}{*}{\#tokens (all)} & Patch & 5{,}746 & 8{,}492 & 11{,}898 & 13{,}086 & 15{,}833 \\
 & Full-gen & 19{,}904 & 23{,}218 & 28{,}695 & 32{,}166 & 59{,}825 \\
\bottomrule
\end{tabular*}
\caption{Patch-based editing vs. full plan generation (GPT-5, single step $T{=}1$, $K{=}5$, top-30 TPC-DS). Patch-based produces a valid plan for $70\%$ of samples, vs. $54\%$ for full generation. \#tokens (out) counts only the emitted patch / plan, while \#tokens (all) counts everything generated including reasoning.}
\label{tab:full_vs_patch}
\vspace{-5mm}
\end{table}

\section{Related Work}

\paragraph{\textbf{Query optimization.}}
Query optimization has long centered on cost-based plan selection over cardinality estimates~\cite{chaudhuri1998overview}, where errors in those estimates are the primary driver of plan-quality degradation~\cite{lohman2014query}. Classical responses include combinatorial join-enumeration algorithms (System R's dynamic programming~\cite{selinger1979access}, DPccp~\cite{moerkotte2006dpccp}, randomized methods~\cite{steinbrunn1997heuristic}) and learned cardinality estimators that approximate selectivities from training data~\cite{kipf2019learned,yang2020neurocard}. The latter, while data-driven, encode per-column or per-join statistics rather than the semantic content of column names, predicates, and values. As a result, neither analytical nor learned estimators directly capture functional dependencies across columns (e.g., ``Model = Accord'' implies ``Make = Honda'') or value-distribution intuitions (e.g., that a single year is selective on a century-scale dimension table). LLMs, in contrast, draw on semantic priors learned from text and code, and prior work shows they can predict such cross-column correlations even from schema names alone~\cite{trummer2023correlations}.

\paragraph{\textbf{Test-time Optimization and Evolutionary Strategies.}}
A growing body of work optimizes individual instances at test time, with methods such as ShinkaEvolve~\cite{lange2025shinkaevolve}, AlphaEvolve~\cite{novikov2025alphaevolve}, TextGrad~\cite{yuksekgonul2025optimizing}, TTT-Discover~\cite{ttt-discover2026}, and GEPA~\cite{agrawal2025gepa} refining candidate solutions against fitness functions. Such per-instance effort is worthwhile when a single instance is itself a high-impact problem, as with a query plan that runs repeatedly at scale. This mirrors a broader move in AI-driven systems research toward using LLMs to refine individual system instances against empirical performance verifiers~\cite{cheng2025barbariansgateaiupending}. We adopt a basic evolutionary search to demonstrate the effectiveness of the optimization framework itself.

\paragraph{\textbf{LLMs for Database Optimization.}}
LLMs have entered query optimization at several stages of the query pipeline. Some rewrite the input SQL before it reaches the optimizer (E3~\cite{xu2025e3}) or restructure the logical plan (Galois~\cite{Satriani2025LogicalAP}). Others act at the optimizer itself, producing the execution plan in its place (LLM-QO~\cite{tan2025can}, LLMOpt~\cite{yao2025query}) or steering its choices with hints (LLM4Hint~\cite{liu2025llm4hint}). A more radical line bypasses the engine altogether, synthesizing executable code from scratch (GenDB~\cite{lao2026gendb}, Bespoke OLAP~\cite{wehrstein2026bespokeolap}). Through a harness, \name{} exposes Apache DataFusion's internal physical plan to an LLM that refines it at test time with small, engine-native edits. Working at the physical-plan level is more concrete than rewriting SQL, enabling execution-level optimizations such as join reordering, and more structured than generating code, keeping the edits interpretable. Because patched plans run natively, the approach reuses the engine's mature optimizations and stays cheap and easy to integrate, whereas synthesizing plans or whole engines rebuilds execution from scratch, a trade-off we quantify in Section~\ref{sec:analysis} (Table~\ref{tab:full_vs_patch}). This suits OLAP workloads, whose repeated heavy queries reward the optimization effort.
\section{Limitations and Future Work}
\label{sec:limitations}

\paragraph{\textbf{Prototypical evaluation.}}
Our evaluation currently focuses on join reordering and join-side selection because cardinality estimation errors are shown to be the dominant source of plan suboptimality~\cite{lohman2014query}. Moreover, the two evolutionary strategies we evaluate, \texttt{PST-Evol} and \texttt{BoL-Evol}, are intentionally lightweight baselines within a modest call budget, and the query suite is restricted to deterministic outputs so that output equivalence serves as a strict, automatable correctness signal. Finally, the scale-transfer study (Section~\ref{sec:scale_transfer}) is demonstrative, and extensions to real deployments would likely require adaptations across multiple aspects besides the scale factor, such as hardware, workload mix, and infrastructure pricing.

However, our harness accommodates substantially broader scope. For instance, alternative optimization targets, such as predicate rewriting, filter placement across UNION ALL branches, and join-strategy changes (e.g., \texttt{NestedLoopJoin} to \texttt{HashJoin}), are expressible through the same patch interface. More advanced test-time optimization strategies, such as self-refinement loops conditioned on execution feedback~\cite{madaan2023selfrefine}, or LLM-driven discovery procedures~\cite{ttt-discover2026}, can be smoothly hosted on our \name{} harness. The LLM's context can also be enriched with the outputs of learned cardinality estimators~\cite{kipf2019learned,yang2020neurocard}, and the LLM itself can be specialized through domain-specific fine-tuning~\cite{tan2025can} to both better adapt to specific workloads and potentially reduce inference cost. Stronger correctness guarantees through constrained edit interfaces, such as a domain-specific patch language that makes invalid plan edits unrepresentable, together with stress tests that mutate the underlying database, are also natural directions for reliable production use.

\paragraph{\textbf{Deployment considerations.}}
An optimized plan binds to the engine's view of the data and schema at search time. When new rows are added under the same distribution, the original optimization continues to apply. Even if the schema or distribution changes, the LLM can be re-prompted with the existing optimization and a diff of what changed to produce an adapted version. Alternatively, the search can be rerun from scratch on the new state. Both routes use our framework's validation and optimization machinery, and the fallback mechanism highlighted also in Algorithm~\ref{alg:evolution} reverts to the baseline plan if no candidate survives validation and provides optimization.

\paragraph{\textbf{Statistics, hardware, and workload context.}}
The same prompt interface that delivers row counts, byte sizes, and column types can also deliver much richer statistics (histograms, distinct counts, \texttt{ANALYZE}-style summaries), hardware specifications such as CPU and memory budgets and parallelism settings, and workload context such as expected query mix, freshness requirements, and data-lake features, for the LLM to reason over at test time. We currently do not exercise these extensions here. Furthermore, while DataFusion has been a well-optimized OLAP engine for years, we could not find a public \texttt{ANALYZE}-equivalent to inject histograms and richer statistics. The relative contributions of compensation for missing statistics and complementary semantic reasoning therefore remain difficult to separate. However, the reasoning trace for the case-study query (Appendix~\ref{sec:cot_attribution}) and the structural-proxy evidence from the join-footprint analysis (Appendix~\ref{sec:join_footprint}) suggest the LLM performs genuine semantic reasoning that can augment static statistics, but a definitive separation is a limitation and left to future work.

\section{Conclusion}

We introduce \name{}, a harness for test-time optimization of physical query plans on Apache DataFusion. \name{} exposes the post-optimizer physical plan to an LLM through a token-efficient serialization and a patch-based editing interface, and we instantiate the workflow with GPT-5 and an evolutionary sampling strategy that uses end-to-end execution time as the fitness signal.

Across LLM-generated query suites on TPC-H and TPC-DS, the framework delivers median speedups of $1.10$--$1.12\times$ and $1.05$--$1.07\times$ respectively, with individual queries reaching $4.78\times$. As OLAP templates are reused repeatedly, these per-query gains amortize the one-time search cost (Appendix~\ref{sec:cost_benefit}). The improvements persist on canonical expert-written TPC and JOB/IMDB queries, transfer from SF-$3$ to SF-$10$ with positively correlated speedups (Section~\ref{sec:scale_transfer}), and hold under an open-weights model (Kimi K2.5), which produces a comparable speedup distribution. The optimizations come from genuine semantic reasoning over schemas and value distributions (Figure~\ref{fig:case_study_viz}, Appendix~\ref{sec:cot_attribution}).

Test-time optimization with LLMs complements classical cost-based optimizers, adding semantic reasoning to their statistics-driven decisions. \name{} is released as a benchmark and harness, and we hope it serves as a base for richer optimization strategies, broader transformation scopes, and tighter correctness guarantees in future work.

\section*{Acknowledgments}
We thank Modal for providing computational credits for this work.

\bibliographystyle{ACM-Reference-Format}
\bibliography{references}

\clearpage
\newpage
\appendix
\section*{Lessons Learned}

We would like to conclude this paper with a set of lessons learned that we think are important for the broader community.

During this project, we explored different methods for timing query execution. Query timing is an inherently stochastic process affected by many factors. When evaluating locally on a single machine, other processes interfere with and affect the evaluation. We tried to minimize disk access noise by using the memory file system for the underlying parquet files corresponding to the dataset tables.

We tested scaling on a single bare metal AWS machine, but this required both CPU pinning and disabling hyper-threading to ensure each query was processed by a single core. Additionally, using a single machine limited our scaling capabilities.

We tested different cloud providers, but none could provide the guarantees that bare metal offers. We eventually settled on Modal, as it allowed us to easily scale execution and reduce noise as much as possible, though not eliminate it. Carefully tuning machine parameters (threads and CPUs) and forcing instances to deploy in the same availability zone helped stabilize execution timing, yielding the CoV values reported in our sensitivity analysis (Section~\ref{apdx:sensitivity}).

\section{Plan Transfer Implementation Details}
\label{app:transfer_details}

To make the transfer described in Section~\ref{sec:scale_transfer} reliable and inexpensive at evaluation time, we implement transfer as a deterministic, rule-based script that is implemented \emph{once} and then reused across queries. Concretely, we create a transfer function consuming: (i) original-scale (e.g., SF3) base plan structure, (ii) original-scale succinct table information, (iii) original-scale optimized plan structure (by applying the optimization patch), (iv) target-scale (e.g., SF10) base plan structure, and (v) target-scale succinct table information. The script then modifies the original-scale optimized plan structure and rewrites it into a runnable target-scale structure.

At a high level, the transfer script follows three principles. First, it \emph{anchors scan nodes} using succinct table metadata by matching scans through a normalized signature (e.g., schema/projection/predicate) that omits dataset-dependent details, yielding a stable mapping from original-scale scan identifiers to target-scale scan identifiers. Second, it \emph{preserves plan intent} by rewriting only identifiers and references while keeping the operator structure unchanged (e.g., join rewiring, join-side swaps, projection updates) as dictated by the original-scale optimization. Third, it enforces \emph{safety checks} (e.g., no dangling references, valid node structure, and immutability checks) before returning a transferred plan.

We note that transferred plans are not necessarily optimal at the target scale: at larger scales, the engine may select different baseline physical operators (e.g., alternative join strategies or repartitioning decisions), and plan choices can depend on scale-sensitive statistics and memory pressure. Nevertheless, our study shows that a deterministic, rule-based transfer mechanism can \emph{reliably replay} many optimization decisions discovered at small scale onto larger-scale plans, preserving correctness while retaining meaningful speedups. This supports a practical workflow for real deployments: discover effective plan transformations cheaply at a prototypical scale, then lift them to production-scale executions by applying principled, automated rewrites, which are particularly valuable for high-stakes large queries. Notably, because the target-scale baseline plan is computed regardless, a transferred plan can be benchmarked against it before deployment and discarded if the performance is not improved, so transfer adds no production risk. Overall, these results motivate future work on more scale-aware transfer policies that better account for target-scale operator selection and resource constraints.

\section{Verification Experiments}
\label{apdx:verification}
\subsection{Sensitivity Analysis for Our Evaluation Protocol}
\label{apdx:sensitivity}
\begin{table}[t]
\centering
\normalsize
\begin{tabular*}{0.9\columnwidth}{@{\extracolsep{\fill}}lcccc}
\toprule
& \multicolumn{2}{c}{TPC-H CoV (\%)} & \multicolumn{2}{c}{TPC-DS CoV (\%)} \\
\cmidrule(lr){2-3}\cmidrule(lr){4-5}
Estimator & p50 & p90 & p50 & p90 \\
\midrule
\texttt{mean}   & 16.00 & 23.44 & 13.06 & 22.14 \\
\texttt{median} &  6.36 &  8.93 &  4.11 &  5.93 \\
\texttt{min}    &  \textbf{2.25} & \textbf{3.13} & \textbf{2.85} & \textbf{4.66} \\
\bottomrule
\end{tabular*}
\caption{Sensitivity analysis of runtime variance under our evaluation protocol.
For each query and estimator (\texttt{mean} / \texttt{median} / \texttt{min}), we run our protocol for $15$ independent waves. At each wave, we collect $R{=}50$ timing repetitions and aggregate them into a single runtime estimate. We then compute CoV (\%) \emph{across waves} per query and report the p50/p90 of these per-query CoVs across the benchmark subset. Min-of-$R$ yields the lowest variability (e.g., TPC-DS p90 CoV $=4.66\%$).}
\label{tab:sensitivity_cov}
\vspace{-8mm}
\end{table}

\label{sec:external_verification}
\begin{figure*}[t]
  \centering
  \includegraphics[width=1.5\columnwidth]{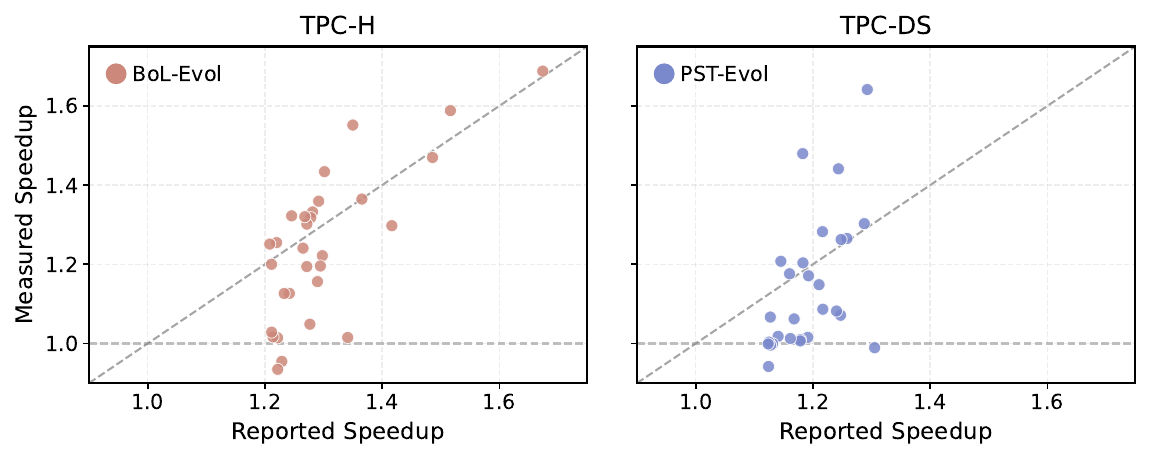}
    \caption{External verification of reported speedups. For each dataset, we re-execute the top-30 optimizations (by reported speedup) in a static, bare-metal environment. Each point compares a query’s reported speedup to its re-measured speedup. We verify that the top reported results largely remain as speedups (most points above $y=1$) and are broadly centered around $y=x$, supporting the validity of our reported gains. Two TPC-DS points are omitted for readability: $(4.78,\, 4.05)$ and $(1.52,\, 2.78)$.
}
  \label{fig:report_vs_measured_topk}
\end{figure*}

\begin{figure*}[t]
  \centering
  \includegraphics[width=0.9\textwidth]{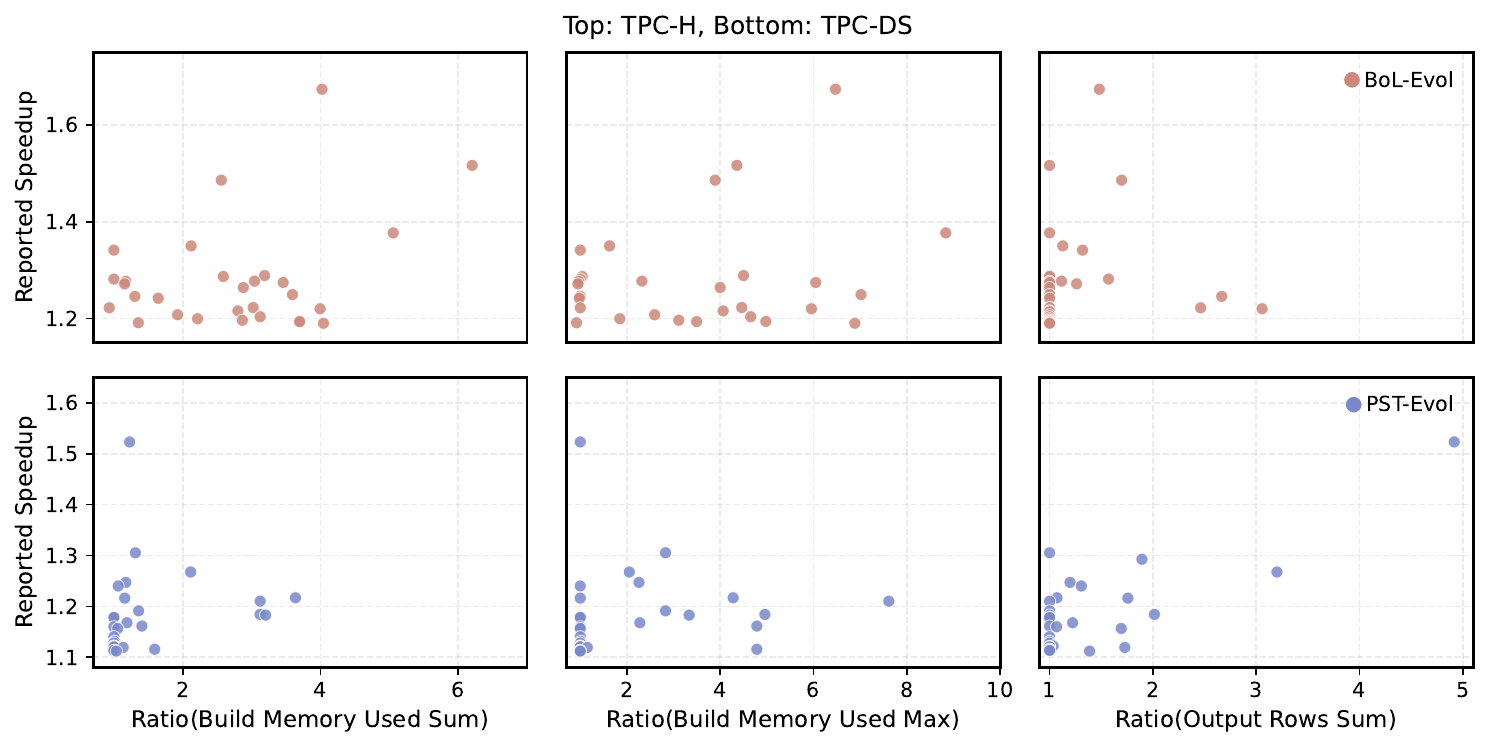}
    \caption{Footprint ratios vs.\ speedup. For each dataset's top-30 optimizations (by speedup), we plot the relationship between speedup and the ratio (base over optimized plan) for complementary \texttt{HashJoin} footprints: (i) output rows (sum), and (ii) the associated build memory used (sum/max). Ratios $>1$ indicate reductions, supporting and helping explain the reported speedups.}
  \label{fig:metric_ratios_vs_speedup}
\end{figure*}

We conduct a sensitivity analysis to calibrate our execution-time estimation protocol, focusing on the estimator choice used to aggregate repeated measurements into a single runtime estimate. We compare three commonly used estimators: \texttt{mean}, \texttt{median}, and \texttt{min}.

Prior to any optimization, we benchmark baseline (unmodified) plans on a subset of queries drawn from each dataset. Specifically, we sample $8$ queries per complexity level, yielding $32$ queries per dataset. For each query, we simulate our execution protocol for $15$ waves and repeat execution $R = 50$ times per wave. Treating each wave as an independent timing estimate, we compute the coefficient of variation (CoV) across waves to assess estimator stability.

As in Table~\ref{tab:sensitivity_cov}, the \texttt{min} estimator exhibits the lowest variance and is the most stable across datasets, with p90 CoV values of $3.13\%$ on TPC-H and $4.66\%$ on TPC-DS. In contrast, the \texttt{mean} estimator is highly unstable in our noisy execution environment, reaching p90 CoV values of approximately $20\%$, while the \texttt{median} offers moderate stability (p90 CoV of $8.93\%$ on TPC-H and $5.93\%$ on TPC-DS).

Importantly, all timings are performed in waves of batched executions ($32$ queries per batch), mirroring the execution structure used in our main experiments. This batching reliably captures platform-level variability and ensures that the sensitivity analysis faithfully reflects the noise characteristics of the full evaluation pipeline.

\begin{table*}[t]
\centering
\small
\setlength{\tabcolsep}{8pt}
\begin{tabular}{llcccccccc}
\toprule
Dataset & Metric & $n$ & P25 & P50 & P75 & P90 & Max & $>\!1\times$ \\
\midrule
\multirow{3}{*}{TPC-H}
  & build-mem sum & 15 & 1.000 & 2.766 & 10.807 & 18.202 & 47.014 & 73.3\% \\
  & build-mem max & 15 & 1.000 & 2.766 & 11.662 & 31.232 & 47.014 & 73.3\% \\
  & output rows   & 15 & 1.000 & 1.000 &  1.040 &  1.665 & $8{,}379$ & 26.7\% \\
\midrule
\multirow{3}{*}{TPC-DS}
  & build-mem sum & 15 & 1.000 & 1.000 & 1.026 & 4.001 & 9.268 & 40.0\% \\
  & build-mem max & 15 & 1.000 & 1.000 & 1.079 & 4.882 & 5.587 & 33.3\% \\
  & output rows   & 15 & 1.000 & 1.000 & 1.000 & 3.521 & 4.004 & 20.0\% \\
\midrule
\multirow{3}{*}{IMDB (JOB)}
  & build-mem sum & 32 & 1.000 & 1.034 & 3.683 & 68.444 & $2{,}470$ & 50.0\% \\
  & build-mem max & 32 & 1.000 & 1.000 & 4.407 & 68.826 & $2{,}524$ & 43.8\% \\
  & output rows   & 32 & 1.000 & 1.000 & 2.376 & 10.502 & 135.833  & 37.5\% \\
\midrule
\multirow{3}{*}{IMDB q-id $16^{*}$}
  & build-mem sum & 4 & 4.785 & 4.988 & 5.570 & 6.503 & 7.125 & 100.0\% \\
  & build-mem max & 4 & 9.000 & 9.069 & 9.181 & 9.381 & 9.515 & 100.0\% \\
  & output rows   & 4 & 1.000 & 1.037 & 2.613 & 5.383 & 7.230 & 50.0\% \\
\bottomrule
\end{tabular}
\caption{Structural-proxy metric ratios (baseline / optimized; $>\!1$ indicates a reduction) on the canonical expert-written queries evaluated in Section~\ref{sec:analysis}, under \texttt{PST-Evol} ($T{=}4$, $K{=}5$). The JOB q-id $16^{*}$ row reports the four variants of structure $16$ run separately on $8$\,GB sandboxes. Maxima for output rows on TPC-H and IMDB are rounded to integers.}
\label{tab:real_world_footprint}
\end{table*}%

\subsection{External Verification on a Stable Environment}
To further support our reported speedups, we select the top 30 queries per dataset (top $25$th percentile) based on the best reported speedup, and re-execute these optimizations in an external and static environment.

We re-evaluate these queries on a bare-metal Linux machine (NixOS, kernel 6.12.62) equipped with an AMD 9900X processor (24 cores, 48 threads), 64 GB of memory, and an NVMe SSD. Each query was executed three times, and the observed run-to-run variation was within 5\%, indicating that the measured speedups are stable and not dominated by transient system effects.

Figure~\ref{fig:report_vs_measured_topk} compares, for each query, the reported speedup (x-axis) against the re-measured speedup (y-axis). Across both datasets, the points concentrate near the $y=x$ line, with deviations split roughly evenly above and below it, suggesting no systematic inflation or deflation of the reported gains. Most queries remain above the $y=1$ baseline under re-measurement, indicating that the optimizations generally preserve their improvements in an external environment; only a small subset falls near or slightly below $1$. For TPC-DS, we omit two points that lie outside the plot’s axis limits for readability; their re-measured speedup coordinates are $(4.78,\, 4.05)$ and $(1.52,\, 2.78)$. Overall, these results validate that the top reported improvements persist under cross-system transfer to a different hardware and OS configuration, even though the exact speedup magnitudes shift between systems.

\subsection{JOIN Footprint Verification}
\label{sec:join_footprint}

Moreover, we conduct a JOIN-footprint analysis to quantify the impact of optimizations through complementary footprints measured during execution. We then examine how changes in these measured quantities relate to the observed speedups. For this, we select the top 30 queries per dataset (top 25th percentile) based on the best reported speedup, and collect footprint statistics for each baseline/optimized pair. We specifically focus on reductions in three metrics by the optimizations:
\begin{itemize}
    \item \emph{Build Memory Used (sum)}: sum of build memory used across all \texttt{HashJoin} operators in the plan (as reported by runtime metrics).
    \item \emph{Build Memory Used (max)}: maximum build memory used among \texttt{HashJoin} operators in the plan.
    \item \emph{Output Rows (sum)}: sum of rows output by \texttt{HashJoin} operators in the plan.
\end{itemize}
In Figure~\ref{fig:metric_ratios_vs_speedup}, we scatter each optimized plan instance where the x-axis is the ratio of the metric between the baseline execution and the optimized execution (baseline / optimized), and the y-axis is the speedup. Across all three metrics, higher speedups are predominantly associated with ratios greater than 1, indicating that improvements in runtime typically coincide with reductions in build-side memory usage and/or join output cardinalities. We observe ratios up to $6\times$ for \emph{Build Memory Used (sum)}, $9\times$ for \emph{Build Memory Used (max)}, and $5\times$ for \emph{Output Rows (sum)}. For presentation, we omit a small number of outliers: for TPC-H, we omit one optimization with speedup $1.21$ achieving $21.5\times$ and $53.7\times$ reductions in \emph{Build Memory Used (sum/max)}; for TPC-DS, we omit two optimizations, including one with speedup $4.78$ achieving $3.3\times$ and $1.6\times$ reductions in \emph{Build Memory Used (sum/max)} and a $1.9\times$ reduction in \emph{Output Rows (sum)}. Overall, with this footprint view, we provide an independent corroboration of the reported speedups.

We extend our footprint analysis to the canonical, expert-written benchmarks evaluated in Section~\ref{sec:analysis} (TPC-H, TPC-DS, and JOB), recording the same three metrics across baseline and \texttt{PST-Evol} ($T{=}4$, $K{=}5$) optimized plan pairs. As reported in Table~\ref{tab:real_world_footprint}, the median ratios generally remain around $1$, but the upper percentiles reveal substantial improvements, with maximum build-memory reductions reaching $47\times$ for TPC-H and $2{,}524\times$ for JOB. Furthermore, the JOB q-id $16^{*}$ variants (executed in isolated $8$\,GB sandboxes) yield consistent $4$--$9\times$ median reductions across both memory metrics. This further corroborates the speedups reported in Section~\ref{sec:analysis}.

\section{Method and Algorithm Details}
\subsection{Complexity Descriptions}
\label{sec:complexity_desc}

To make the query-generation prompt precise without overconstraining style, we use a coarse \emph{complexity} label during generation, helping us specify the intended structural ingredients (e.g., join patterns, aggregation, and nesting such as subqueries/CTEs/windows), for producing a workload with controlled diversity and comparable difficulty across queries. Below we include the exact rubric used to instantiate these complexity levels (which is part of the query generation prompt for \name \; in Appendix~\ref{sec:query_gen_prompts})

\begin{schemabox}{Complexity Descriptions}
{\footnotesize

\textbf{Complexity 1-2:}
\begin{verbatim}
Simple queries with basic SELECT statements, a join and simple 
WHERE clauses
\end{verbatim}
\tcbline

\textbf{Complexity 3-4:}
\begin{verbatim}
SELECT with WHERE clause and 2-3 JOINs, start adding aggregations
\end{verbatim}
\tcbline

\textbf{Complexity 5-6:}
\begin{verbatim}
Even more JOINs between tables, GROUP BY, HAVING clauses, basic 
subqueries
\end{verbatim}
\tcbline

\textbf{Complexity 7-8:}
\begin{verbatim}
Complex JOINs, window functions, CTEs, nested subqueries
\end{verbatim}
\tcbline

\textbf{Complexity 9-10:}
\begin{verbatim}
Highly complex queries with multiple nested subqueries, advanced 
window functions, CTEs, and intricate JOINs
\end{verbatim}
}
\end{schemabox}

\subsection{Pseudocode of the Evolutionary Algorithms}
We share the pseudocode of our evolutionary strategies in Algorithm~\ref{alg:evolution}.
\begin{algorithm}[h]
\caption{Evolutionary Sampling for Query Plan Optimization (\texttt{PST-Evol} and \texttt{BoL-Evol})}
\label{alg:evolution}
\begin{algorithmic}[1]
\REQUIRE Query set $\mathcal{Q}$; baseline planner $\textsc{Plan}_0(\cdot)$; LLM optimizer;
         iterations $T$; population size $K$; mode $\in \{\textsc{PST},\textsc{BoL}\}$
\ENSURE Sampled plans $\{\hat p_t^{(k)}\}$ with validity $v_t^{(k)}$ and timings $\tau_t^{(k)}$; $\forall k \in [K]$

\FOR{each $q \in \mathcal{Q}$}
    \STATE $p_0 \leftarrow \textsc{Plan}_0(q)$
    \STATE $p_0^{(k)} \leftarrow p_0$ \COMMENT{initialize threads}

    \FOR{$t = 1$ to $T$}
        \STATE $\Delta_t^{(k)} \leftarrow \textsc{LLM}(q, p_{t-1}^{(k)})$ \COMMENT{sample an optimization patch}
        \STATE $\hat p_t^{(k)} \leftarrow \textsc{Apply}(p_{t-1}^{(k)}, \Delta_t^{(k)})$ \COMMENT{apply the patch and modify the plan}
        \STATE $(v_t^{(k)}, \tau_t^{(k)}) \leftarrow \textsc{Validate\&Time}(\hat p_t^{(k)}, p_0)$
        \COMMENT{$v_t^{(k)}{=}1$ iff successfully executes \& output matches baseline}

        \IF{mode == \textsc{PST}}
            \STATE $p_t^{(k)} \leftarrow \begin{cases}
                \hat p_t^{(k)} & \text{if } v_t^{(k)}=1 \\
                p_{t-1}^{(k)}  & \text{otherwise}
            \end{cases}$ \COMMENT{valid advances, else keep parent}
        \ELSE
            \STATE $k^\star \leftarrow \arg\min_{k:\, v_t^{(k)}=1}\; \tau_t^{(k)}$
            \STATE $p_t^{(k)} \leftarrow \begin{cases}
                \hat p_t^{(k^\star)} & \text{if } \exists k: v_t^{(k)}=1 \\
                p_{t-1}^{(k)}        & \text{otherwise}
            \end{cases}$ \COMMENT {broadcast best-valid if exists}
        \ENDIF
    \ENDFOR
\ENDFOR
\end{algorithmic}
\end{algorithm}

\subsection{Convergence and Comparison with \texttt{Best-of}}
\label{sec:best_of_n}
\begin{figure*}[t]
  \centering
  \includegraphics[width=\linewidth]{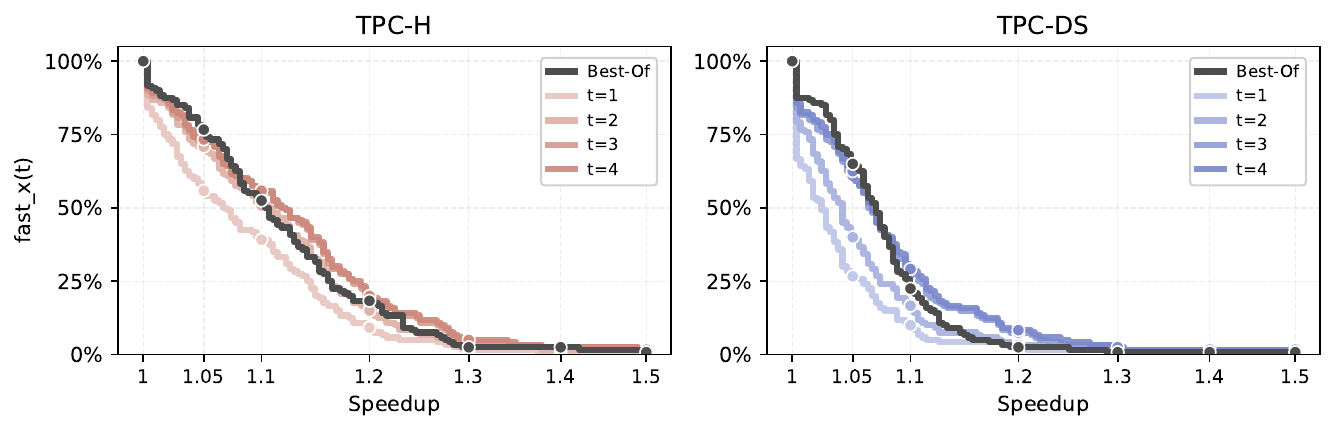}
  \caption{Linear Y-axis Version of the CCDF Plots in Figure~\ref{fig:ccdf_sep}. Here, we can more clearly see that roughly top 40-60\% percentile of the frontier speedups are thanks to the evolutionary algorithms.}
  \label{fig:ccdf_sep_lin}
\end{figure*}

\begin{figure*}[t]
  \centering
  \includegraphics[width=\linewidth]{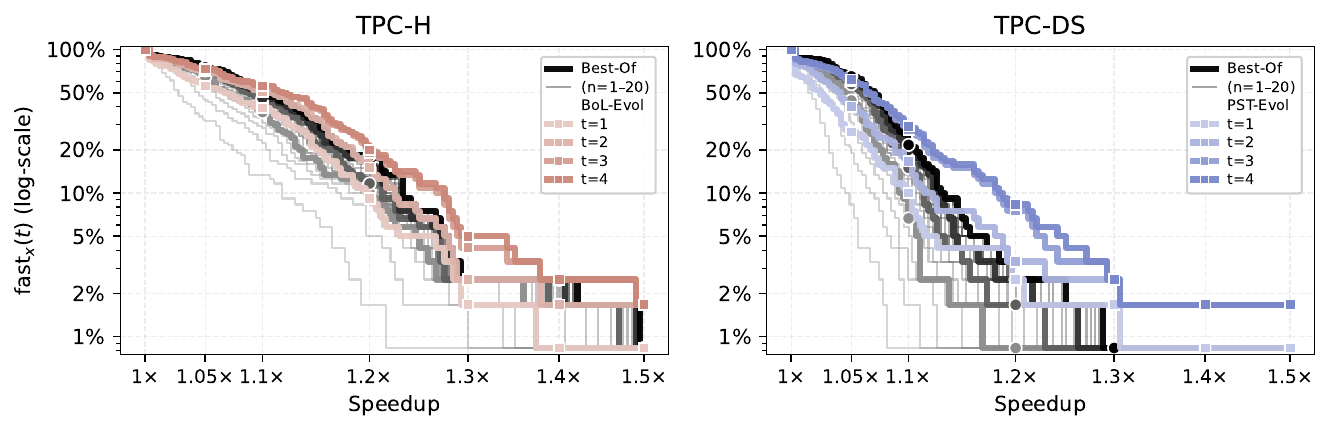}
  \caption{Matched-budget comparison of \texttt{Best-of} against the evolutionary strategies. Thin gray curves show the \texttt{Best-of} CCDF at each sample count $n{=}1,\dots,20$ (thick at $n{=}5,10,15,20$); colored curves show the per-dataset evolutionary strategy at each step ($T\times K$ samples, $K{=}5$ per step). \texttt{Best-of} saturates beyond $n\approx10$, while the evolutionary strategies push further into the upper tail, most visibly for \texttt{PST-Evol} on TPC-DS.}
  \label{fig:best_of_n}
\end{figure*}

We examine how the evolutionary algorithms converge and how they compare to \texttt{Best-of} under an equal budget of $T\times K$ samples ($K{=}5$ per step, up to $20$ at $T{=}4$). Figure~\ref{fig:ccdf_sep_lin} re-plots Figure~\ref{fig:ccdf_sep} on a linear $y$-axis, where it is clearer that \texttt{Best-of} leads at the lower percentiles, covering roughly the bottom $40$--$60\%$ of speedups (up to about $1.1\times$). Figure~\ref{fig:best_of_n} then overlays the \texttt{Best-of} frontier at every sample count $n{=}1,\dots,20$ against the per-dataset evolutionary strategy at each step (\texttt{BoL-Evol} for TPC-H, \texttt{PST-Evol} for TPC-DS). We observe that the evolutionary algorithms push the upper tail beyond what \texttt{Best-of} reaches even at $n{=}20$, most visibly on TPC-DS, where \texttt{PST-Evol} raises the P90 from $1.127$ to $1.183$ and the P95 from $1.157$ to $1.241$. The evolutionary algorithms are also more sample-efficient. On TPC-H, \texttt{BoL-Evol} reaches the median speedup that \texttt{Best-of} attains at $20$ samples using only $15$ (a $1.33\times$ saving), while on TPC-DS the two are comparable. \texttt{Best-of} concentrates its improvement in its first samples and changes little beyond $n\approx10$. The two strategies are thus complementary, with \texttt{Best-of} stronger at the lower percentiles and evolution at the upper percentiles.

\section{Correlation Analyses}
\subsection{Complexity Analysis}
\begin{figure*}[t]
  \centering
  \includegraphics[width=0.9\linewidth]{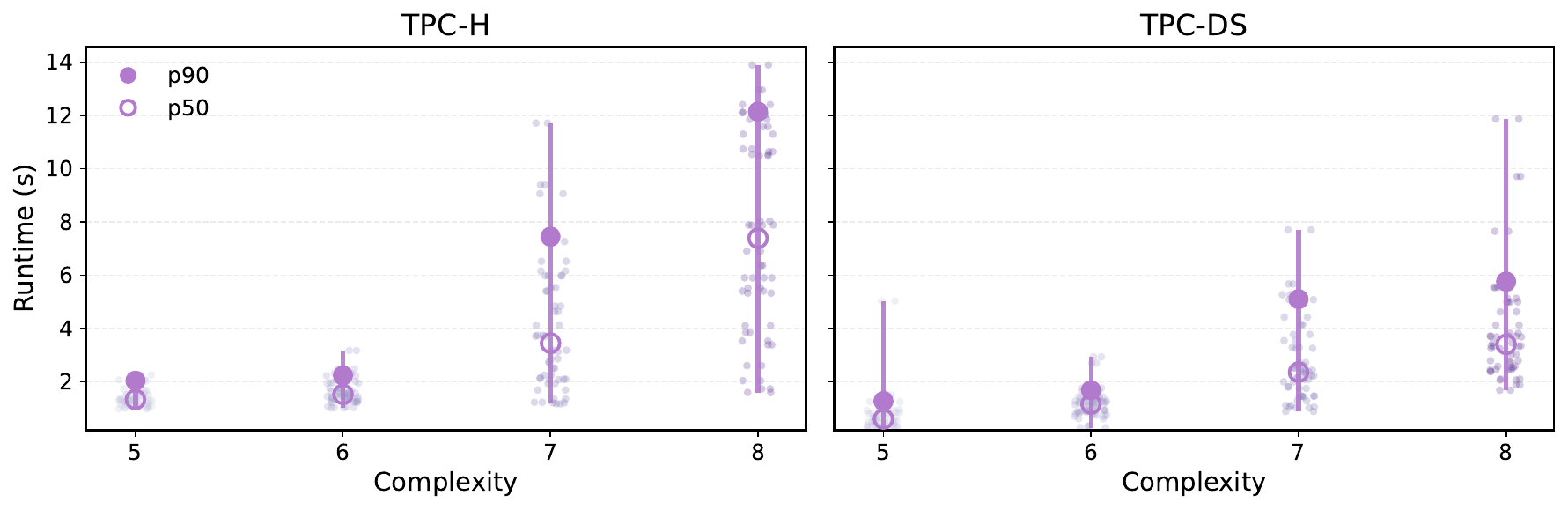}
  \caption{Complexity vs. Runtime. We observe an increasing trend: the higher the complexity, the longer a query runs.}
  \label{fig:comp_vs_runtime}
\end{figure*}

\paragraph{\textbf{Complexity vs. Runtime.}} Figure~\ref{fig:comp_vs_runtime} presents the distribution of execution runtime for queries of different complexities. We observe that there is a positive relationship: more complex queries take longer to complete.

\begin{figure*}[t]
  \centering
  \includegraphics[width=\linewidth]{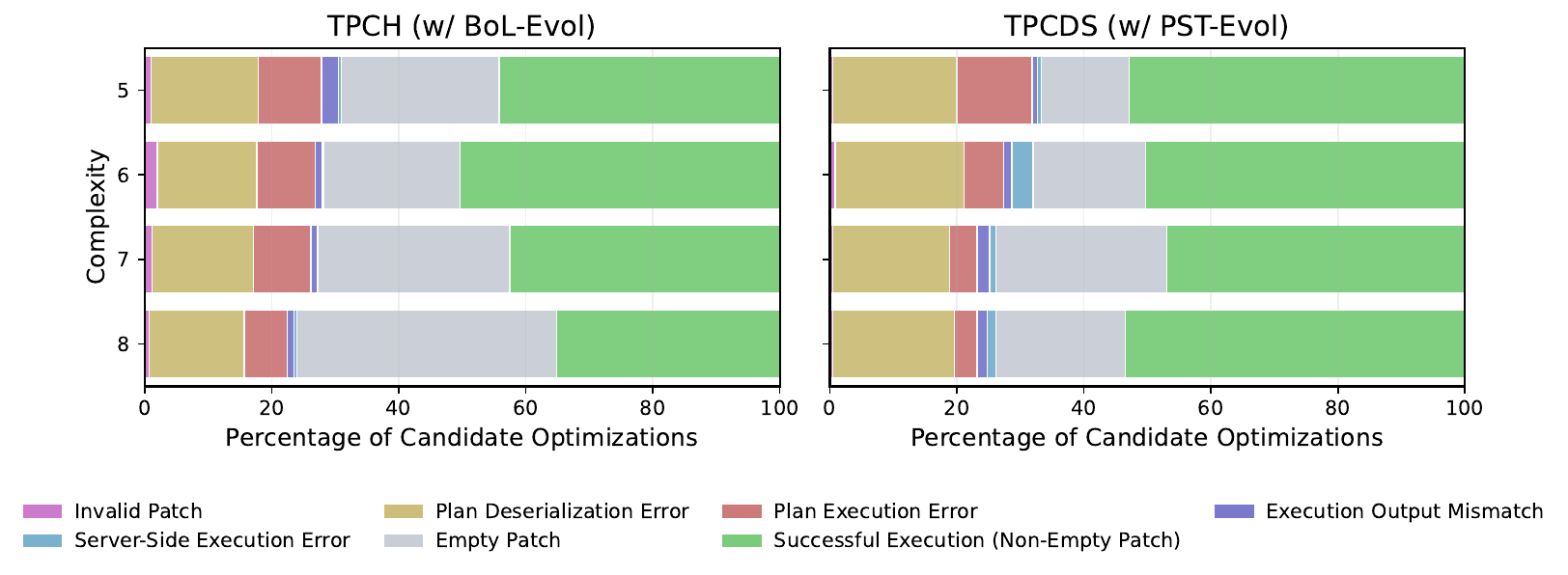}
  \caption{Complexity vs.\ optimization outcomes. Stacked bars show the breakdown of candidate outcomes across query complexities for TPC-H (BoL-Evol) and TPC-DS (PST-Evol).}
  \label{fig:comp_vs_outcomes}
\end{figure*}

\paragraph{\textbf{Complexity vs. Optimization Outcomes.}}
Figure~\ref{fig:comp_vs_outcomes} breaks down the outcomes of candidate optimizations produced during evolutionary sampling, grouped by query complexity. For each generated candidate, we assign the outcome based on the first outcome mode encountered in our pipeline:
\begin{itemize}
    \item \emph{Invalid Patch}: The JSON Patch cannot be parsed or applied to the serialized plan.
    \item \emph{Plan Deserialization Error}: The patched JSON plan cannot be converted back into a valid DataFusion execution plan.
    \item \emph{Plan Execution Error}: The plan is successfully deserialized but throws an error during execution.
    \item \emph{Execution Output Mismatch}: The plan executes but produces results that do not match the baseline query output. Thus the optimized plan is semantically not equivalent to the baseline query plan, therefore invalid.
    \item \emph{Server-side Execution Error}: Rare infrastructure failures in the sandboxed environment.
    \item \emph{Empty Patch}: The model explicitly returns no edits to the original plan (i.e., no optimizations found).
    \item \emph{Successful Execution}: The model returns non-empty patches which make the plan execute successfully and match the baseline output.
\end{itemize}
Across both datasets, we observe that the LLM rarely produces syntactically invalid patches, indicating that structured patch generation is reliable even when operating over large, nested plan representations. The dominant failure modes instead arise after patch application: a good fraction of patched plans either cannot be deserialized into DataFusion’s internal execution representation, or fail mid-execution. While less frequent, we also observe occasional output mismatches, suggesting that semantic drift is possible even when edits appear structurally valid; these cases are filtered out by our output-equivalence check and treated as unsuccessful candidates. Furthermore, we observe a shift in the model’s behavior that with increasing complexity, the overall failure rates tend to decrease, but \emph{empty patch} rates increase (particularly for complexities 7--8 relative to 5--6) in both datasets, which may indicate that the model becomes more conservative on larger and more intricate plans. Overall, $40\%-50\%$ of the model's generations are non-empty and execute successfully.

\begin{figure*}[h]
  \centering
  \includegraphics[width=0.9\linewidth]{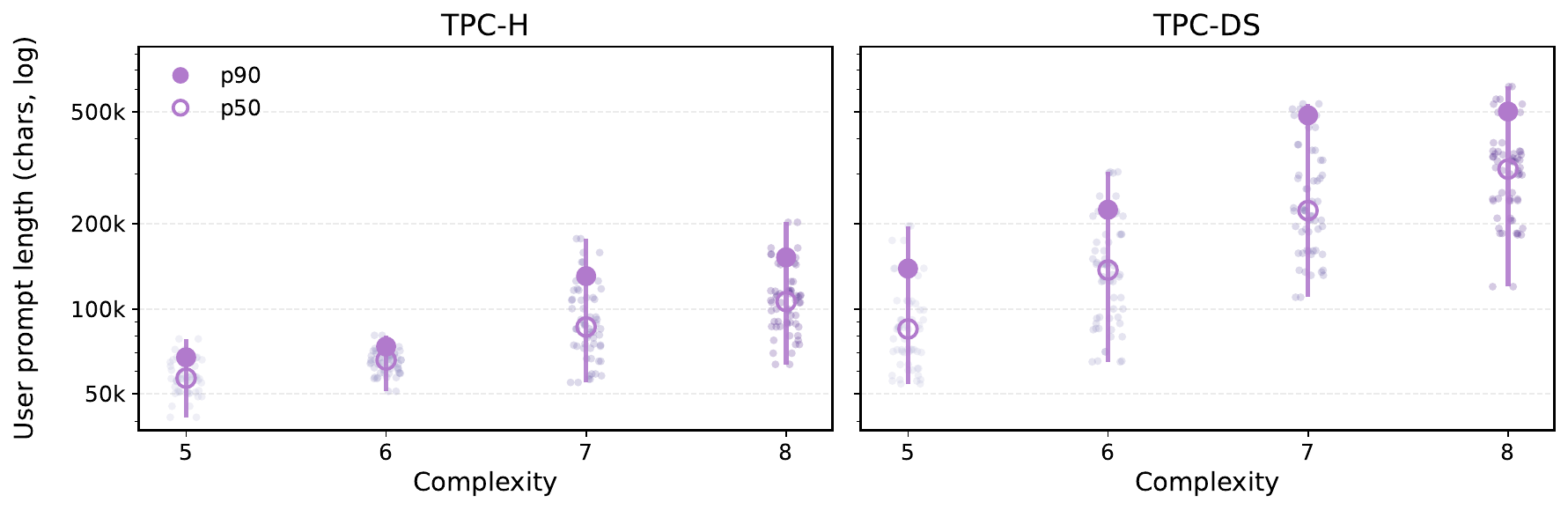}
  \caption{Complexity vs. User Prompt Length. We observe an increasing trend for the length of the User Prompt employed during optimization of the physical plans. The format could be seen in Appendix~\ref{prompt:opt_user}. The length of prompts for the TPC-DS is notably longer than those for TPC-H.}
  \label{fig:comp_vs_length}
\end{figure*}

\textbf{Complexity vs. Prompt Lengths.} \label{sec:comp_vs_speedup} Figure~\ref{fig:comp_vs_length} illustrates the relationship between the complexities and the \emph{User Prompt} length. We observe a positive relationship, as the more complex the query, the longer the user prompt becomes. Moreover, we observe notably longer prompts for the TPC-DS dataset, reaching up to 500k characters. Table~\ref{tab:prompt_component_breakdown} further shows the breakdown of the \emph{User Prompt} into its components. We observe that the majority of the \emph{User Prompt} comes from the structural representation of the physical plan. Moreover, the table information (presented succinctly) represents a notable portion as well. Comparatively, the \emph{System Prompt} is only about $3-9\%$ of the \emph{User Prompt}. Finally, regarding the compression of the full table information (mentioned in Section~\ref{sec:dbplanbench}), presenting the information succinctly helps compress about $20\times$ (for a median query), indicating that such compression helps the prompt length not blow up, as otherwise the table information would make the prompt $5-10\times$ longer, causing context length limitations.

\subsection{Speedup vs. Query Characteristics}
\label{sec:speedup_vs_char}
\begin{figure*}[t]
  \centering
  \includegraphics[width=0.9\linewidth]{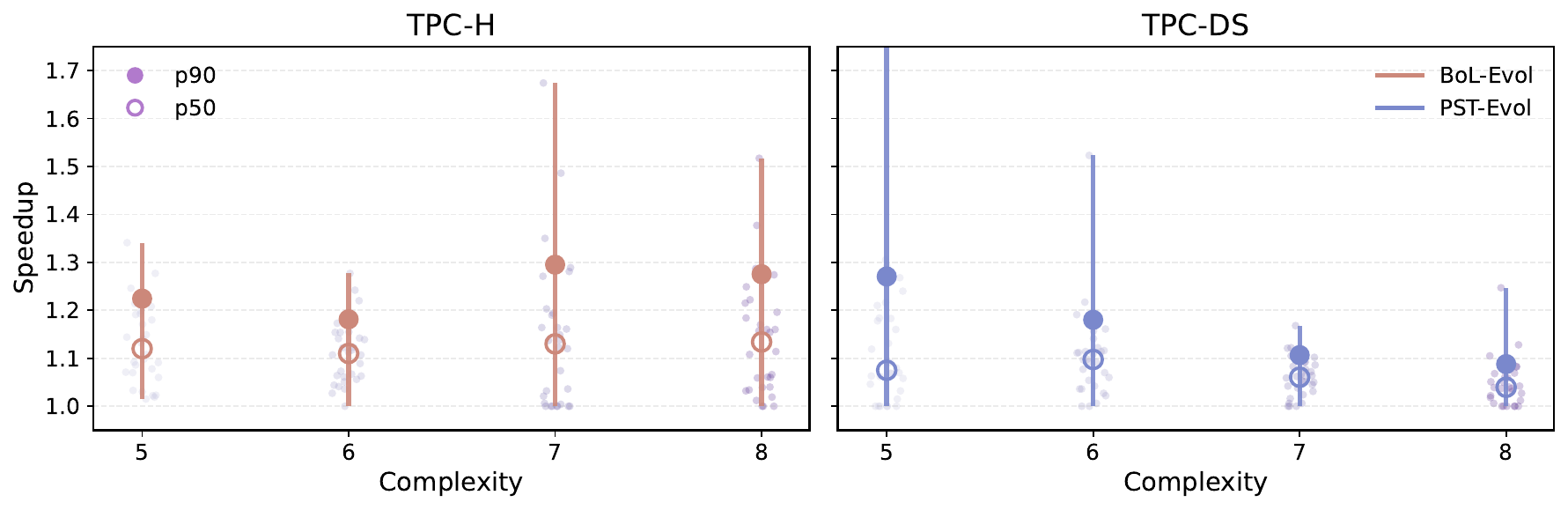}
  \caption{Complexity vs. Speedup. We observe that in TPC-H, there is a slight increase in speedup with respect to complexity, whereas in TPC-DS, there is a visible inverse correlation.}
  \label{fig:comp_vs_speedup}
\end{figure*}

\begin{figure}[h]
  \centering
  \includegraphics[width=\linewidth]{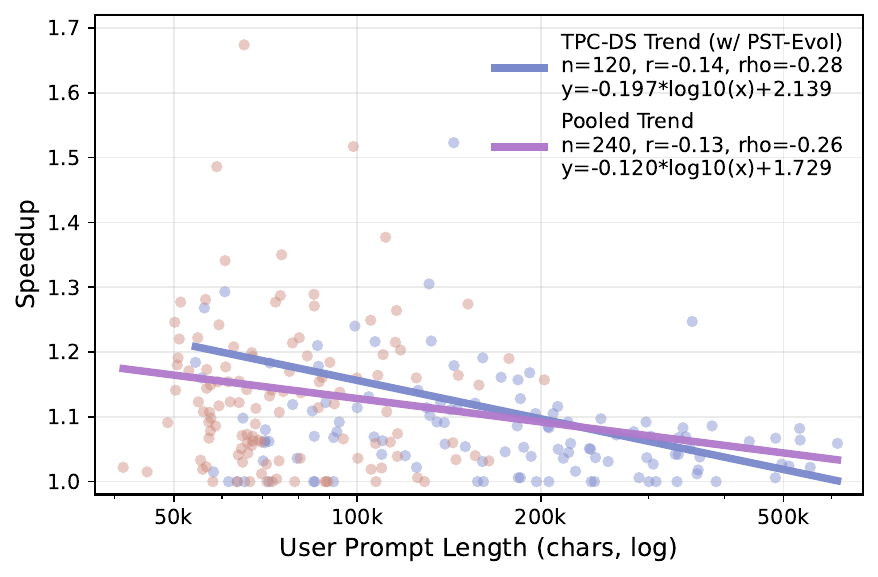}
  \caption{Prompt length vs.\ speedup.
Scatter plot of user prompt length (characters; log-scale) versus achieved speedup, with fitted linear trends for TPC-DS and pooled data (TPC-DS: Pearson $r{=}-0.14$, Spearman $\rho{=}-0.275$; pooled: $r{=}-0.13$, $\rho{=}-0.26$), indicating a \emph{weak negative association} between prompt length and achieved speedup.}
  \label{fig:prompt_vs_speedup}
\end{figure}

\paragraph{\textbf{Complexity vs.\ Speedup.}}
Building on the above analyses (complexity vs.\ runtime, outcomes, and prompt lengths), we examine whether more complex queries yield larger speedups. Figure~\ref{fig:comp_vs_speedup} shows that for TPC-H, speedups increase slightly with complexity, whereas for TPC-DS we observe a clear inverse correlation. This behavior is consistent with prompt length: Figure~\ref{fig:prompt_vs_speedup} reports a weak negative association between user prompt length and speedup (TPC-DS: Pearson $r=-0.14$, Spearman $\rho=-0.275$; pooled: $r=-0.13$, $\rho=-0.26$). Overall, query complexity alone is not a reliable predictor of optimization gains, as realized speedups may be dependent on whether a query's semantics and plan structure admit improvements under our validity constraints.

\begin{figure*}[t]
  \centering
  \includegraphics[width=\linewidth]{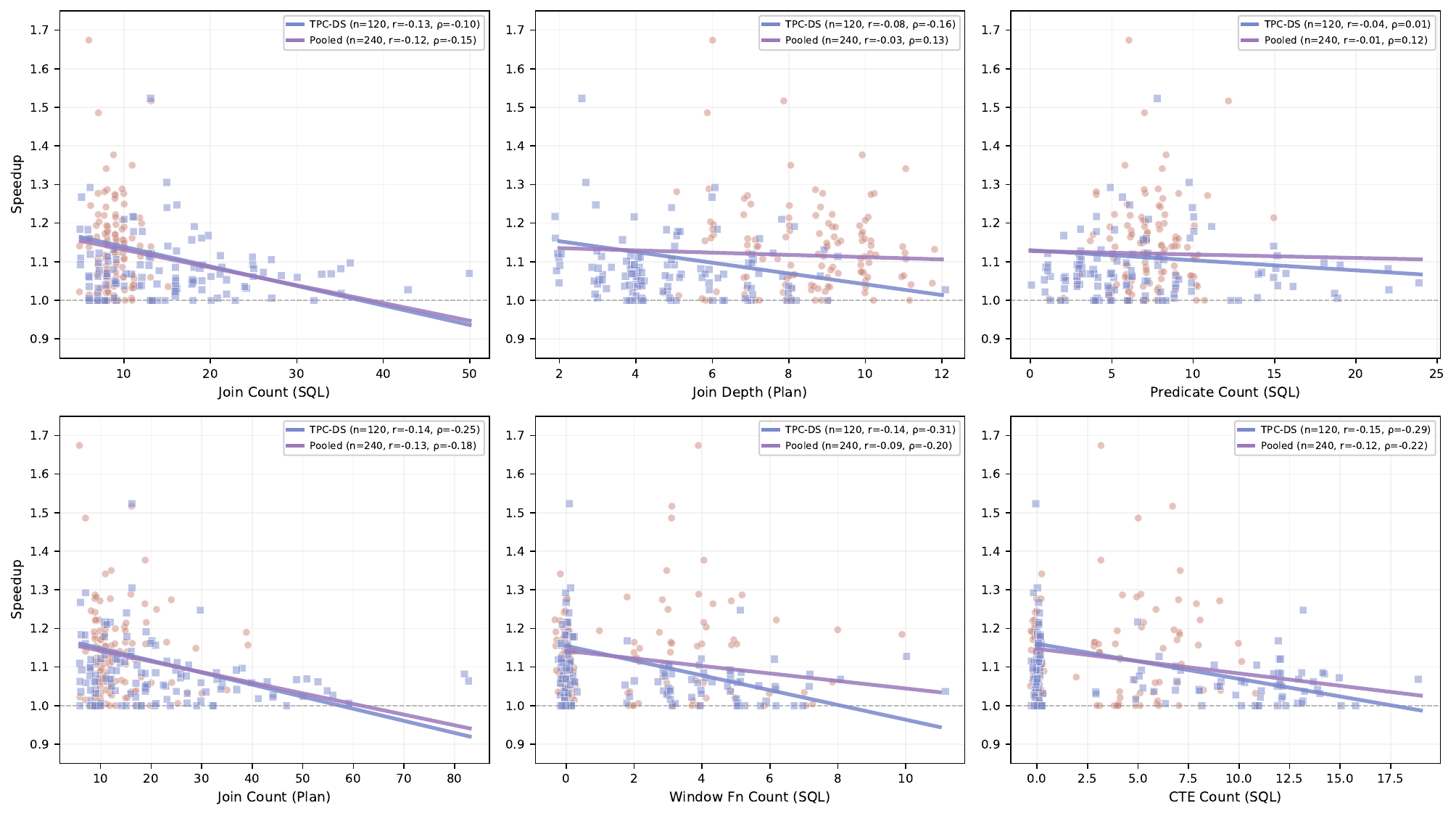}
  \caption{Speedup vs.\ query characteristics. Each panel scatters per-query speedup against one characteristic, with TPC-DS and pooled linear trends and their Pearson $r$ and Spearman $\rho$ in the legend. Across all six panels the correlations are weak, with $r$ within $[-0.15,-0.01]$ and $\rho$ within $[-0.31,0.13]$, indicating a weak, generally negative tendency.}
  \label{fig:speedup_vs_char}
\end{figure*}

\paragraph{\textbf{Finer Characteristics vs.\ Speedup.}}
We further examine whether finer query characteristics predict the achieved speedup, using six features drawn from two sources. From the SQL text, we count joins (\texttt{JOIN} keywords), predicates (atomic \texttt{WHERE}/\texttt{HAVING} conditions), window functions (\texttt{OVER} clauses), and CTEs (\texttt{WITH} definitions). From the physical plan, we measure the join count (the number of join operators) and the join depth (the longest chain of joins from root to leaf). Figure~\ref{fig:speedup_vs_char} plots speedup against each. We observe a weak negative tendency overall, both pooled and on TPC-DS, meaning that queries scoring higher on these features tend to achieve slightly smaller speedups. The tendency is clearest for window-function count (pooled $r{=}-0.09$, $\rho{=}-0.20$; TPC-DS $r{=}-0.14$, $\rho{=}-0.31$) and CTE count (pooled $r{=}-0.12$, $\rho{=}-0.22$; TPC-DS $r{=}-0.15$, $\rho{=}-0.29$), while the remaining four features show weaker and less consistent associations. As with query complexity, we observe that no single characteristic reliably predicts how much a query can be sped up.

\begin{table*}[t]
\centering
\small
\setlength{\tabcolsep}{8pt}

\begin{tabular}{
  l
  c c c
  c
  c c c
}
\toprule
Dataset
& \multicolumn{3}{c}{User Prompt Composition (p50 \%)}
& System Prompt
& \multicolumn{3}{c}{Table Info Compression Rate} \\
\cmidrule(lr){2-4}\cmidrule(lr){5-5}\cmidrule(lr){6-8}
& Query
& Structure
& \shortstack[c]{Succinct\\Table Info}
& \shortstack[c]{p50 \% of\\User Prompt}
& Min
& p50
& Max \\
\midrule
TPC-H
& 2.10 & 65.40 & 32.10
& 8.40
& 14.27$\times$ & 18.20$\times$ & 21.93$\times$ \\
TPC-DS
& 2.10 & 51.60 & 46.20
& 3.40
& 14.17$\times$ & 21.05$\times$ & 28.00$\times$ \\
\bottomrule
\end{tabular}
\caption{Prompt composition and table-information compression by dataset. First block: median (p50) percentage breakdown of the \emph{user prompt} into query text, structural information, and succinct table information (Appendix~\ref{prompt:opt_user}). \emph{System Prompt} reports the fixed system-prompt length (5{,}995 chars; Appendix~\ref{prompt:opt_system}) as a percentage of user-prompt length. Final block: min/p50/max compression rate when converting full table info to the succinct representation provided to LLM (full table info is not provided at runtime). Percentages are per-metric medians (p50) computed independently across queries and may not sum to 100\%.}
\vspace{-3mm}

\label{tab:prompt_component_breakdown}
\end{table*}

\section{Optimization Time Amortization}
\label{sec:cost_benefit}

Although the search time for an optimization can exceed the execution time itself, OLAP workloads where templates are reused at scale amortize this upfront cost across many subsequent executions~\cite{van2024tpc,10825377}. We focus on optimization time as our primary cost metric, as wall-clock time is directly measurable in our evaluations and maps cleanly onto the operational goals of a deployment, such as faster response times and higher throughput on analytical workloads. The figures reported below reflect our evaluation setup, particularly for optimizing single queries in isolation at SF-$3$.

\subsection{Measurement Protocol}

To quantify this, we evaluate two query sets from TPC-DS, a stratified random sample of $32$ queries (complexity levels $5$--$8$) and the top-30 queries ranked by reported \texttt{PST-Evol} ($T{=}4$, $K{=}5$) speedup. We instantiate our protocol with \texttt{PST-Evol}, as in the main experiments. At each step of the search, we record two primary latencies for each evaluated candidate. First, we track the wall-clock latency of completing $K$ GPT-5 API calls in parallel, denoted $\tau_{\text{api}}$. Second, we benchmark the candidate's execution plan concurrently across $R{=}50$ Modal sandboxes (each provisioned with $4$ CPUs and $4$\,GB of RAM), yielding a set of execution times $\{t_1, \ldots, t_{50}\}$. Similarly, because the $R$ sandboxes execute in parallel, the evaluation phase completes only when the slowest instance finishes, so we model the wall-clock time of one Modal evaluation step with the maximum runtime, $\tau_{\text{modal}} = \max_i t_i$.

As the optimization process generates and evaluates $K$ candidates in parallel over $T$ sequential steps, the total search time is bottlenecked by the sequential transitions, which lead to the total optimization time $D_{\text{time}}$,
\[
D_{\text{time}} \;=\; T\,\bigl(\tau_{\text{api}} + \tau_{\text{modal}}\bigr).
\]
With the total search time established, we define the crossover threshold $n^\star$ as the number of production executions required for the cumulative time savings to overtake the upfront search time. Using the expected runtime $\mu = \tfrac{1}{R}\sum_i t_i$ on the unoptimized baseline plan to characterize one production execution, the break-even point is
\[
n^\star \;=\; \frac{D_{\text{time}}}{\mu\,(1 - 1/s)},
\]
where $s$ is the speedup obtained for that specific query.

\subsection{Break-Even Analysis}

Table~\ref{tab:cb_budget} details the optimization time budget across both query sets. The median end-to-end optimization requires roughly $9$ to $12$ minutes. For the top-30 set, the time crossover occurs at approximately $2{,}000$ executions (dropping below $1{,}500$ in the lower quartile), while the random sample reaches crossover at roughly $3{,}250$ executions. These thresholds fall within reuse counts of production OLAP templates. For instance, on the top-billed tail of the Redshift fleet, $300$ instances processed $27{,}441{,}359$ user-executed queries over $15$ days, roughly $6{,}000$ per instance per day, of which $61.8\%$ repeat previously seen queries~\cite{wu2024stage}. Thus, the recurring templates accumulate well beyond our crossover thresholds, so the framework pays for its search time quickly.

Several deployment dynamics may push this crossover point earlier in practice. Faithfully transferring an optimization from a smaller scale factor to a larger one (Section~\ref{sec:scale_transfer}) increases the expected execution time $\mu$, lowering the break-even count proportionally. Alternatively, the dominant search component is $\tau_{\text{api}}$, and any reduction in $\tau_{\text{api}}$ (with the speedup $s$ held fixed) would carry through to $D_{\text{time}}$, whether from faster inference, prompt caching for repeated schema context, or domain-specific fine-tuning (e.g., LLM-QO~\cite{tan2025can}).

Referring back to $D_{\text{time}} = T(\tau_{\text{api}} + \tau_{\text{modal}})$, the $\tau_{\text{modal}}$ term scales with the baseline runtime $\mu$. Assuming the API latency $\tau_{\text{api}}$ does not depend on $\mu$, the crossover correlates inversely with $\mu$ via the API term, suggesting that the framework is most economical for queries with longer baseline runtimes.

Our analysis is prototypical and anchored to a specific evaluation setup on Modal sandboxes. While we treat time savings as a proxy for operational gains such as faster query execution and higher pipeline throughput, translating them into a net monetary gain depends on deployment-specific realities. Infrastructure pricing, request batching, token pricing, and the local business value of faster execution would all vary too widely for a generalized metric. We treat $D_{\text{time}}$ and $n^\star$ as the cleanest, most objective proxies our framework can offer, and encourage future work to conduct a deployment-grade economic study.

\section{Anatomy of the Optimization}
\label{sec:anatomy}

\subsection{Reasoning-Trace Attribution for the Top Query}
\label{sec:cot_attribution}
To attribute the case-study optimizations (Section~\ref{case-study}) to the model's own reasoning, we reproduce the same optimization for the TPC-DS query achieving a $4.78\times$ of speedup, by modifying the prompt to include a request for the model to output its reasoning before the optimization patch. For each step of the optimization, we record this reasoning and the resulting patch, evaluated with our standard protocol ($R{=}50$, min estimator, correctness checks). The excerpts below are verbatim spans of the model's output, abridged with \texttt{[...]} and with the key parts in bold.

\begin{table*}[t]\centering\small
\setlength{\tabcolsep}{8pt}
\begin{tabular}{lrrrrrrrr}
\toprule
 & \multicolumn{4}{c}{Random ($n=32$)} & \multicolumn{4}{c}{Top-30} \\
\cmidrule(lr){2-5}\cmidrule(lr){6-9}
Metric & P25 & P50 & P75 & P90 & P25 & P50 & P75 & P90 \\
\midrule
API latency $\tau_{\text{api}}$ (s) & $93.8$ & $115$ & $158$ & $195$ & $107$ & $160$ & $219$ & $256$ \\
Modal eval $\tau_{\text{modal}}$ (s) & $11.4$ & $15.9$ & $30.8$ & $34.2$ & $10.2$ & $13.2$ & $20.3$ & $25.0$ \\
Total opt.\ time $D_{\text{time}}$ (s) & $438$ & $554$ & $731$ & $815$ & $500$ & $690$ & $925$ & $1{,}080$ \\
Crossover $n^\star$ (reps) & $1{,}769$ & $3{,}253$ & $6{,}423$ & $10{,}494$ & $1{,}371$ & $1{,}980$ & $4{,}122$ & $6{,}280$ \\
\bottomrule
\end{tabular}
\caption{Optimization time budget per query on TPC-DS. Total optimization time follows $D_{\text{time}} = T\,(\tau_{\text{api}} + \tau_{\text{modal}})$ with $T{=}4$, $K{=}5$, and the crossover follows $n^\star = D_{\text{time}}/(\mu\,(1 - 1/s))$, where $s$ is the speedup from \texttt{PST-Evol}, under the same evaluation setup.}
\label{tab:cb_budget}
\end{table*}

\textbf{Build-side selection (step~1).} The model first estimates the base-table sizes and observes that the joins to the small \texttt{item} and \texttt{customer\_address} dimensions build their hash tables on the much larger combined sales stream rather than on the dimensions themselves, and proposes swapping the two sides.
\begin{tracebox}{Reasoning excerpt ($T{=}1$)}\small
\textbf{Cardinality estimates (bottom-up).} \texttt{date\_dim}: 73{,}049 rows; filter \texttt{d\_year=2001} $\rightarrow$ \textbf{$\approx$365 rows (1 year of days)}; \texttt{item}: 36{,}000 rows (dimension); \texttt{customer\_address}: 94{,}000 rows (dimension). UNION ALL of channels $\approx$15.12M rows.\elision
Hash joins should build on the smaller side to minimize memory and shuffle. Join to \texttt{item} (node 37) builds on the large 15M fact side (left) and probes with \texttt{item} (right). \textbf{This is suboptimal; flip to build on item}. Join to \texttt{customer\_address} (node 33) builds on the large fact+item side (left) and probes with address (right). Also suboptimal; flip to build on address.\elision
Building hash tables on small dimensions (36k items and 94k addresses) instead of multi-million-row facts dramatically reduces memory use and shuffle volumes. Given our estimates, we \textbf{reduce build size by $\sim$400$\times$ (15M $\rightarrow$ 36k)} at node 37 and by $\sim$150$\times$ (15M $\rightarrow$ 94k) at node 33, without changing output schemas.
\end{tracebox}
We observe that this swap lowers hash-join build memory by $3.3\times$ (sum) and $1.6\times$ (max).

\textbf{Date-before-address reordering (step~2).} The model next turns to join order. It reasons that filtering to a single year is highly selective, so the \texttt{date\_dim} join should run before the expensive \texttt{customer\_address} join.
\begin{tracebox}{Reasoning excerpt ($T{=}2$)}\small
\textbf{Pushing the selective \texttt{date\_dim} join (with \texttt{d\_year=2001}) as early as possible} reduces the row volume flowing into downstream repartitions/joins, especially the expensive hash repartition on \texttt{addr\_sk} for the \texttt{customer\_address} join.\elision
Current plan joins \texttt{date\_dim} (365 rows) AFTER both item and customer\_address joins; that misses an opportunity to prune early. Improved: Join \texttt{date\_dim} right after item join (i.e., before \texttt{customer\_address}) to \textbf{prune $\sim$80\% rows early (assuming sales distributed across $\sim$5 years; 2001 $\approx$ 1/5)}.\elision
After: date\_dim join \textbf{cuts fact stream to $\sim$3.02M ($\approx$20\% of 15.12M)}, then we repartition/join by \texttt{addr\_sk}. This reduces shuffle volume and hash table sizes for the \texttt{customer\_address} join by $\sim$80\%.
\end{tracebox}
We observe that applying the year filter earlier yields a $1.30\times$ reduction in the summed hash-join output rows.

\textbf{Date-before-item reordering (step~3).} The model then pushes the \texttt{date\_dim} join earlier still, ahead of the \texttt{item} join, so the large shuffle on \texttt{item\_sk} runs on the already-filtered stream.
\begin{tracebox}{Reasoning excerpt ($T{=}3$)}\small
\textbf{The most selective operation is the date filter} via the join with \texttt{date\_dim} (to year 2001). In the current plan, the enormous repartition by \texttt{item\_sk} and hash join with \texttt{item} (node 39$\rightarrow$37) happen before the date filter (node 26). This means the engine shuffles and joins $\sim$15.1M rows by \texttt{item\_sk}, then reduces to $\sim$3M on the date join.\elision
If we reorder to apply the date join before the item join, we only repartition $\sim$3M rows by \texttt{item\_sk}, \textbf{around $5\times$ less shuffle and join work}. This is a substantial cost reduction and preserves semantics because the item and date joins are independent (different keys: \texttt{item\_sk} vs \texttt{sold\_date\_sk}).\elision
TPC-DS facts are concentrated in \textbf{$\sim$1998--2002; 2001 covers roughly 20\% of those years' rows}.
\end{tracebox}
We observe this further reduces the summed hash-join output rows from $1.30\times$ to $1.90\times$.

\textbf{No further edit (step~4).} At the last step the model looks for further reorderings but finds none that both reduce cost and stay valid.
\begin{tracebox}{Reasoning excerpt ($T{=}4$)}\small
The plan already exhibits good join-side selection for all hash joins and pushes the year filter into the scan. There is no semantics-preserving structural modification (swapping sides or reordering joins) that both reduces cost and satisfies the constraints \textbf{without introducing cycles or forcing a larger left input}.\elision
this is a critical case where \textbf{no structural JSON patch is advisable}.
\end{tracebox}
The model applies no further patch, and the chain ends. Its three patches together cut total build memory by $3.3\times$, peak per-task build memory by $1.6\times$, and the summed hash-join output rows by $1.9\times$, reproducing the case study's $4.78\times$ speedup.

Across the four steps, the reasoning rests on knowledge that the plan statistics do not contain, chiefly that a single calendar year is a small slice of a multi-year fact table. The model turns this judgment into the join-reordering and build-side edits that \name{} targets, which we observe translate into fewer intermediate rows, lower build memory, and ultimately the speedup. It also stops once no valid edit remains, rather than degrading the plan.

\subsection{Plan Representation and Optimization Patch}
\label{sec:exec_plan_details}
We examine the optimization from our case study (Section~\ref{case-study}), the actual plan that \texttt{PST-Evol} ($T{=}4$, $K{=}5$) found for the TPC-DS query with the $4.78\times$ speedup. We present its serialized physical plan and the resulting patch. This is the main-experiment optimization, distinct from the reproduction behind the reasoning trace in Appendix~\ref{sec:cot_attribution}.

To enable LLMs to reason about physical query plans, we first serialize the in-memory physical plan into a JSON representation.
However, this raw serialization has two drawbacks. It is extremely verbose, embedding detailed scan metadata, statistics, and type information that are repeated across partitions, and it is deeply nested, with all subplans inlined recursively.
While suitable for machine processing, this representation is poorly aligned with the reasoning capabilities of LLMs.
We therefore apply two transformations. First, we flatten the plan into a node list and introduce a \textit{structure} field that captures only the plan topology, explicitly encoding how operators are connected without embedding their full definitions.
This abstraction exposes the join graph and execution shape in a form that is easier for an LLM to analyze. Second, we construct a \textit{succinct\_table\_info} that summarizes operator-level semantics while removing redundant fields and normalizing repeated metadata, yielding a compact and semantically faithful description of the plan. A short example of the schema for the \textit{succinct\_table\_info} and the \textit{structure} fields is shared in the optimization prompts (Appendix~\ref{sec:plan_optim_prompts}); the actual instances for this query are shown below.

The listing below presents three representative nodes from the \textit{structure} field of this query, namely the top projection (node 0), the hashJoin between \texttt{date\_dim} and the sales stream (node 26), and the filter applying \texttt{d\_year=2001} (node 29). The remaining nodes are omitted, marked by \texttt{// ...} comments.

\begin{jsonblock}{\texttt{structure} excerpt for the $4.78\times$ query}
{
  "0": {
    "projection": {
      "expr": [
        {"column": {"name": "sales_year"}},
        {"column": {"index": 1, "name": "channel"}},
        {"column": {"index": 2, "name": "ca_state"}},
        {"column": {"index": 3, "name": "i_category"}},
        {"column": {"index": 4, "name": "total_sales"}},
        {"column": {"index": 5, "name": "total_returns"}},
        {"column": {"index": 6, "name": "net_sales"}}
      ],
      "exprName": [
        "sales_year", "channel", "ca_state", "i_category",
        "total_sales", "total_returns", "net_sales"
      ],
      "input": 1
    }
  },

  // ...

  "26": {
    "hashJoin": {
      "left": 27,
      "on": [
        {
          "left":  {"column": {"name": "d_date_sk"}},
          "right": {"column": {"index": 1, "name": "sold_date_sk"}}
        }
      ],
      "projection": [1, 2, 4, 5, 6, 7],
      "right": 32
    }
  },

  // ...

  "29": {
    "filter": {
      "defaultFilterSelectivity": 20,
      "expr": {
        "binaryExpr": {
          "l": {"column": {"index": 1, "name": "d_year"}},
          "op": "Eq",
          "r": {"literal": {"int64Value": "2001"}}
        }
      },
      "input": 30
    }
  }
  
  // ...
}
\end{jsonblock}

Each entry of the \textit{succinct\_table\_info} field describes a single parquet-scan node and includes its column names and types, the table's total row count, and its total byte size. The LLM does not receive histograms, minimum or maximum values, or distinct counts, and semantic reasoning is necessary to optimize the execution. The listing below presents two representative entries, namely \texttt{date\_dim}, on which the year predicate is applied, and \texttt{store\_sales}, the largest fact table, with approximately 8.6 million rows. The \texttt{id} field of each entry identifies the corresponding scan node within the \textit{structure} field above.

\begin{jsonblock}{\texttt{succinct\_table\_info} excerpt for the $4.78\times$ query}
// Showing date_dim and store_sales. 8 other tables omitted: 
// warehouse, store_returns, web_returns, web_sales, 
// catalog_returns, catalog_sales, item, customer_address.
[
  {
    "id": 31,
    "node": {
      "parquetScan": {
        "baseConf": {
          "constraints": {},
          "fileGroups": [{ "files": [{
            "path": "tmp/data/data_tpcds/date_dim.parquet",
            "size": "1814672",
            "statistics": {
              "n_rows": 73049, "total_bytes": 4006445
            }
          }] }],
          "projection": [0, 6],
          "schema": { "columns": [
            ["d_date_sk", "INT64", true],
            ["d_date_id", "UTF8", true],
            // ... (24 more columns)
            ["d_current_quarter", "UTF8", true],
            ["d_current_year", "UTF8", true]
          ] }
        },
        "predicate": {
          "binaryExpr": {
            "l": {"column": {"index": 6, "name": "d_year"}},
            "op": "Eq",
            "r": {"literal": {"int64Value": "2001"}}
          }
        }
      }
    }
  },

  // ...

  {
    "id": 53,
    "node": {
      "parquetScan": {
        "baseConf": {
          "constraints": {},
          "fileGroups": [{ "files": [{
            "path": "tmp/data/data_tpcds/store_sales.parquet",
            "size": "423638358",
            "statistics": {
              "n_rows": 8639377, "total_bytes": 486965414
            }
          }] }],
          "projection": [0, 2, 6, 9, 20],
          "schema": { "columns": [
            ["ss_sold_date_sk", "INT64", true],
            ["ss_sold_time_sk", "INT64", true],
            ["ss_item_sk", "INT64", true]
            // ... (20 more columns)
          ] }
        }
      }
    }
  }
]
\end{jsonblock}

\captionsetup[subfigure]{justification=centering,singlelinecheck=false}

\begin{figure*}[t]
\centering

\begin{subfigure}[t]{0.49\linewidth}
\centering
\resizebox{\linewidth}{!}{
\begin{tikzpicture}[x=1.0cm,y=1.0cm]
\node[joinop] (J26) at (0,3.0) {HashJoin (26)\\date $\bowtie$ sales};
\node[leaf]  (D29) at (-2.2,1.6) {date\_dim\\$d\_year=2001$\\(29)};
\node[joinop] (J33) at (2.2,1.6) {HashJoin (33)\\+ address};
\node[joinop] (J37) at (1.2,0.5) {HashJoin (37)\\+ item};

\node[leaf]  (U40) at (0.0,-0.8) {SalesUnion\\(40)};
\node[leaf]  (I80) at (2.4,-0.8) {item\\(80)};
\node[leaf]  (A84) at (3.8,0.5) {customer\_address\\(84)};

\draw[build] (D29) -- (J26);
\draw[probe] (J33) -- (J26);

\draw[build] (J37) -- (J33);
\draw[probe] (A84) -- (J33);

\draw[build] (U40) -- (J37);
\draw[probe] (I80) -- (J37);
\begin{scope}[on background layer]
  \draw[rounded corners=6pt, line width=0.6pt, gray!55]
    ([xshift=-4pt,yshift=-4pt]current bounding box.south west) rectangle
    ([xshift= 4pt,yshift= 4pt]current bounding box.north east);
\end{scope}
\end{tikzpicture}
}
\caption{\textbf{Before} (baseline)}
\end{subfigure}
\hfill
\begin{subfigure}[t]{0.49\linewidth}
\centering
\resizebox{\linewidth}{!}{
\begin{tikzpicture}[x=1.0cm,y=1.0cm]
\node[joinop,changed] (J33b) at (0,3.0) {HashJoin (33)\\+ address};
\node[leaf,changed]  (A84b) at (-2.2,1.6) {customer\_address\\(84)};
\node[joinop,changed] (J37b) at (2.2,1.6) {HashJoin (37)\\+ item};
\node[leaf,changed]  (I80b) at (1.2,0.5) {item\\(80)};
\node[joinop,changed] (J26b) at (3.8,0.5) {HashJoin (26)\\date $\bowtie$ sales};

\node[leaf]  (D29b) at (2.6,-0.8) {date\_dim\\$d\_year=2001$\\(29)};
\node[leaf]  (U40b) at (5.0,-0.8) {SalesUnion\\(40)};

\draw[build] (D29b) -- (J26b);
\draw[probe] (U40b) -- (J26b);

\draw[build] (I80b) -- (J37b);
\draw[probe] (J26b) -- (J37b);

\draw[build] (A84b) -- (J33b);
\draw[probe] (J37b) -- (J33b);
\begin{scope}[on background layer]
  \draw[rounded corners=6pt, line width=0.6pt, gray!55]
    ([xshift=-4pt,yshift=-4pt]current bounding box.south west) rectangle
    ([xshift= 4pt,yshift= 4pt]current bounding box.north east);
\end{scope}
\end{tikzpicture}
}

\caption{\textbf{After} (patched)}
\label{fig:join-graph}
\end{subfigure}

\vspace{3mm}

\footnotesize
\noindent
\begin{minipage}[t]{0.49\linewidth}
\textbf{Join order:} date join is applied last.\\
\textbf{Build sides:} build uses large intermediates (SalesUnion at (37), then intermediate at (33)).
\end{minipage}
\hfill
\begin{minipage}[t]{0.49\linewidth}
\textbf{Join order:} date join moved earlier (filters sales before other joins).\\
\textbf{Build sides:} build uses small dimensions (item at (37), address at (33)).
\end{minipage}

\caption{Join-graph digest for the top query in TPC-DS. Node IDs in parentheses match those in the \texttt{structure} excerpt.}
\label{fig:tpcds-top-join-graph}
\end{figure*}

The LLM is then provided with the \textit{structure} and \textit{succinct\_table\_info} and tasked with producing a patch that describes a plan transformation. The patch is a concise, machine-applicable diff (RFC-6902 JSON Patch) over the physical plan, expressing how operators are rewired rather than restating the entire plan. It captures join reordering by reconnecting operators in the join graph while preserving query semantics. To ensure correctness, the patch also updates associated join inputs, join predicates, output projections, and the inputs of feeder operators such as \texttt{coalesceBatches} and \texttt{repartition} so that operator interfaces remain consistent after reordering. This patch-based representation isolates the core optimization (such as applying selective joins earlier to reduce intermediate sizes) while avoiding the verbosity and redundancy of full before/after plan serializations.

The listing below presents the full patch produced for this query after $T=4$ evolutionary steps, with operations that span six nodes.

\begin{jsonblock}{Final aggregated patch for the $4.78\times$ query}
[
  {"op": "replace", "path": "/37/hashJoin/left",       "value": 80},
  {"op": "replace", "path": "/37/hashJoin/right",      "value": 38},
  {"op": "replace", "path": "/37/hashJoin/on/0/left",
    "value": {"column": {"index": 0, "name": "i_item_sk"}}},
  {"op": "replace", "path": "/37/hashJoin/on/0/right",
    "value": {"column": {"index": 2, "name": "item_sk"}}},
  {"op": "replace", "path": "/37/hashJoin/projection", "value": [2, 3, 5, 6, 7, 1]},
  {"op": "replace", "path": "/33/hashJoin/left",       "value": 84},
  {"op": "replace", "path": "/33/hashJoin/right",      "value": 34},
  {"op": "replace", "path": "/33/hashJoin/on/0/left",
    "value": {"column": {"index": 0, "name": "ca_address_sk"}}},
  {"op": "replace", "path": "/33/hashJoin/on/0/right",
    "value": {"column": {"index": 2, "name": "addr_sk"}}},
  {"op": "replace", "path": "/33/hashJoin/projection", "value": [2, 3, 5, 6, 7, 1]},
  {"op": "replace", "path": "/32/coalesceBatches/input", "value": 40},
  {"op": "replace", "path": "/26/hashJoin/projection", "value": [2, 3, 4, 5, 6, 7, 1]},
  {"op": "replace", "path": "/39/repartition/input",   "value": 26},
  {"op": "replace", "path": "/37/hashJoin/projection", "value": [2, 3, 5, 6, 7, 1, 8]},
  {"op": "replace", "path": "/33/hashJoin/projection", "value": [8, 2, 5, 6, 7, 1]},
  {"op": "replace", "path": "/25/coalesceBatches/input", "value": 33}
]
\end{jsonblock}

\subsection{Effect of the Patch on the Join Graph}

As shown in Figure~\ref{fig:tpcds-top-join-graph}, the TPC-DS workload has three sales fact tables (store\_sales, catalog\_sales, web\_sales) that are first combined into a single logical fact stream (“SalesUnion”), and that stream is then joined with small dimension tables to apply predicates and attach descriptive attributes.
It joins the sales facts to date\_dim to enforce a time predicate (d\_year = 2001), then joins to item to restrict and annotate by product attributes, and joins to customer\_address to restrict and annotate by customer location attributes.
Progressive filtering of a large fact table by selective dimension predicates, followed by downstream aggregation and projection, is a common pattern in TPC-DS.

In the baseline (“Before”), the join order delays the date\_dim join until last. The engine first hashes and joins the large SalesUnion with item, then joins that intermediate with customer\_address, and only then applies the date\_dim filter via the final “date $\bowtie$ sales” hash join.
This is costly because the hash-build sides are large intermediates. The first join builds on (or materializes) the large SalesUnion, and the second join builds on an already-expanded intermediate, so both memory footprint and build/probe work are inflated before the year predicate can reduce cardinality.
The patch changes only the physical join graph by pushing the date join earlier so that the d\_year = 2001 predicate filters the sales stream before it participates in other joins;
as a result (“After”), the plan performs the “date $\bowtie$ sales” join first, and subsequent joins to item and customer\_address operate on a much smaller probe input while using small dimensions as hash-build inputs, which reduces build-side size, intermediate cardinalities, and overall join cost.

\clearpage

\section{Data and Prompts}
\subsection{Example Queries}
\subsubsection{Queries from TPC-H}
\begin{tpchquery}{A Query with Complexity 5}
SELECT
  cr.r_name AS region_name,
  DATE_PART('year', o.o_orderdate) AS order_year,
  p.p_brand AS brand,
  SUM(l.l_extendedprice * (1 - l.l_discount)) AS total_revenue,
  SUM(l.l_quantity) AS total_quantity,
  AVG(ps.ps_supplycost) AS avg_supply_cost
FROM orders AS o
JOIN customer AS c ON c.c_custkey = o.o_custkey
JOIN nation AS cn ON cn.n_nationkey = c.c_nationkey
JOIN region AS cr ON cr.r_regionkey = cn.n_regionkey
JOIN lineitem AS l ON l.l_orderkey = o.o_orderkey
JOIN supplier AS s ON s.s_suppkey = l.l_suppkey
JOIN nation AS sn ON sn.n_nationkey = s.s_nationkey
JOIN region AS sr ON sr.r_regionkey = sn.n_regionkey
JOIN part AS p ON p.p_partkey = l.l_partkey
JOIN partsupp AS ps ON ps.ps_partkey = l.l_partkey AND ps.ps_suppkey = l.l_suppkey
WHERE
  o.o_orderdate >= DATE '1994-01-01'
  AND o.o_orderdate < DATE '1996-01-01'
  AND cr.r_regionkey = sr.r_regionkey
GROUP BY
  cr.r_name,
  DATE_PART('year', o.o_orderdate),
  p.p_brand
HAVING
  SUM(l.l_quantity) > (
    SELECT AVG(li2.l_quantity)
    FROM lineitem AS li2
    JOIN part AS p2 ON p2.p_partkey = li2.l_partkey
    WHERE p2.p_brand = p.p_brand
  )
\end{tpchquery}

\begin{tpchquery}{A Query with Complexity 8}
WITH supp AS (
  SELECT
    s.s_suppkey AS supplierkey,
    s.s_name AS s_name,
    n.n_name AS s_nation,
    r.r_name AS s_region
  FROM supplier s
  JOIN nation n ON s.s_nationkey = n.n_nationkey
  JOIN region r ON n.n_regionkey = r.r_regionkey
),
cust AS (
  SELECT
    c.c_custkey AS custkey,
    c.c_name AS c_name,
    n.n_name AS c_nation,
    r.r_name AS c_region
  FROM customer c
  JOIN nation n ON c.c_nationkey = n.n_nationkey
  JOIN region r ON n.n_regionkey = r.r_regionkey
),
base AS (
  SELECT
    s.supplierkey,
    s.s_name,
    s.s_region,
    c.c_region,
    EXTRACT(YEAR FROM o.o_orderdate)::int AS order_year,
    CASE
      WHEN p.p_size <= 10 THEN 'S'
      WHEN p.p_size <= 20 THEN 'M'
      ELSE 'L'
    END AS size_bucket,
    (l.l_extendedprice * (1 - l.l_discount)) AS revenue,
    (ps.ps_supplycost * l.l_quantity) AS cost,
    (l.l_extendedprice * (1 - l.l_discount)) - (ps.ps_supplycost * l.l_quantity) AS margin
  FROM lineitem l
  JOIN orders o ON l.l_orderkey = o.o_orderkey
  JOIN part p ON l.l_partkey = p.p_partkey
  JOIN partsupp ps ON l.l_partkey = ps.ps_partkey AND l.l_suppkey = ps.ps_suppkey
  JOIN supp s ON l.l_suppkey = s.supplierkey
  JOIN cust c ON o.o_custkey = c.custkey
  WHERE
    o.o_orderdate >= DATE '1995-01-01'
    AND o.o_orderdate < DATE '1997-01-01'
    AND l.l_returnflag <> 'R'
    AND p.p_type NOT LIKE '%PROMO%'
),
t_cust_regs AS (
  SELECT
    b.supplierkey,
    b.s_name,
    b.s_region,
    b.order_year,
    b.c_region
  FROM base b
  GROUP BY
    b.supplierkey, b.s_name, b.s_region, b.order_year, b.c_region
),
t_cust_reg_counts AS (
  SELECT
    t.supplierkey,
    t.s_name,
    t.s_region,
    t.order_year,
    COUNT(*) AS cust_region_count
  FROM t_cust_regs t
  GROUP BY
    t.supplierkey, t.s_name, t.s_region, t.order_year
),
t2_main AS (
  SELECT
    b.supplierkey,
    b.s_name,
    b.s_region,
    b.order_year,
    SUM(b.margin) AS sum_margin,
    SUM(b.revenue) AS sum_revenue,
    SUM(b.cost) AS sum_cost,
    SUM(CASE WHEN b.size_bucket = 'S' THEN b.margin ELSE 0 END) AS margin_small,
    SUM(CASE WHEN b.size_bucket = 'M' THEN b.margin ELSE 0 END) AS margin_medium,
    SUM(CASE WHEN b.size_bucket = 'L' THEN b.margin ELSE 0 END) AS margin_large
  FROM base b
  GROUP BY
    b.supplierkey, b.s_name, b.s_region, b.order_year
),
t2 AS (
  SELECT
    m.supplierkey,
    m.s_name,
    m.s_region,
    m.order_year,
    m.sum_margin,
    m.sum_revenue,
    m.sum_cost,
    m.margin_small,
    m.margin_medium,
    m.margin_large,
    COALESCE(c.cust_region_count, 0) AS cust_region_count
  FROM t2_main m
  LEFT JOIN t_cust_reg_counts c
    ON m.supplierkey = c.supplierkey
   AND m.s_name = c.s_name
   AND m.s_region = c.s_region
   AND m.order_year = c.order_year
),
t3 AS (
  SELECT
    t2.*,
    SUM(t2.sum_margin) OVER (PARTITION BY t2.s_region, t2.order_year) AS region_total_margin,
    AVG(t2.sum_margin) OVER (PARTITION BY t2.s_region, t2.order_year) AS region_avg_supplier_margin,
    DENSE_RANK() OVER (PARTITION BY t2.s_region, t2.order_year ORDER BY t2.sum_margin DESC) AS supplier_rank,
    SUM(t2.sum_margin) OVER (
      PARTITION BY t2.s_region, t2.order_year
      ORDER BY t2.sum_margin DESC
      ROWS BETWEEN UNBOUNDED PRECEDING AND CURRENT ROW
    ) AS running_margin,
    LAG(t2.sum_margin) OVER (PARTITION BY t2.supplierkey, t2.s_region ORDER BY t2.order_year) AS prev_year_margin
  FROM t2
)
SELECT
  t3.s_region,
  t3.order_year,
  t3.s_name AS supplier_name,
  t3.sum_margin,
  t3.sum_revenue,
  t3.sum_cost,
  ROUND(t3.sum_margin / NULLIF(t3.sum_revenue, 0), 6) AS margin_rate,
  t3.cust_region_count,
  t3.supplier_rank,
  ROUND(t3.sum_margin / NULLIF(t3.region_total_margin, 0), 6) AS supplier_share_of_region,
  ROUND(t3.running_margin / NULLIF(t3.region_total_margin, 0), 6) AS cumulative_share_in_region,
  (t3.sum_margin - COALESCE(t3.prev_year_margin, 0)) AS yoy_margin_change,
  (SELECT AVG(t2b.sum_margin) FROM t2 t2b WHERE t2b.supplierkey = t3.supplierkey AND t2b.s_region = t3.s_region) AS avg_margin_supplier_region
FROM t3
WHERE
  t3.supplier_rank <= 3
  AND t3.sum_revenue > 0
  AND t3.cust_region_count >= 2
ORDER BY
  t3.s_region,
  t3.order_year,
  t3.supplier_rank
\end{tpchquery}

\subsubsection{Queries from TPC-DS}
\begin{tpcdsquery}{A Query with Complexity 5}
SELECT
  d.d_year AS sales_year,
  s.channel,
  ca.ca_state,
  i.i_category,
  SUM(s.sales_amt) AS total_sales,
  COALESCE(SUM(s.return_amt), 0) AS total_returns,
  SUM(s.sales_amt) - COALESCE(SUM(s.return_amt), 0) AS net_sales
FROM
(
  SELECT
    'store'::text AS channel,
    ss.ss_sold_date_sk AS sold_date_sk,
    ss.ss_item_sk AS item_sk,
    ss.ss_addr_sk AS addr_sk,
    ss.ss_net_paid AS sales_amt,
    sr_ret.return_amt AS return_amt
  FROM store_sales ss
  LEFT JOIN (
    SELECT sr_item_sk, sr_ticket_number, SUM(sr_return_amt) AS return_amt
    FROM store_returns
    GROUP BY sr_item_sk, sr_ticket_number
  ) sr_ret
    ON sr_ret.sr_item_sk = ss.ss_item_sk
   AND sr_ret.sr_ticket_number = ss.ss_ticket_number

  UNION ALL

  SELECT
    'web'::text AS channel,
    ws.ws_sold_date_sk AS sold_date_sk,
    ws.ws_item_sk AS item_sk,
    ws.ws_bill_addr_sk AS addr_sk,
    ws.ws_net_paid AS sales_amt,
    wr_ret.return_amt AS return_amt
  FROM web_sales ws
  LEFT JOIN (
    SELECT wr_item_sk, wr_order_number, SUM(wr_return_amt) AS return_amt
    FROM web_returns
    GROUP BY wr_item_sk, wr_order_number
  ) wr_ret
    ON wr_ret.wr_item_sk = ws.ws_item_sk
   AND wr_ret.wr_order_number = ws.ws_order_number

  UNION ALL

  SELECT
    'catalog'::text AS channel,
    cs.cs_sold_date_sk AS sold_date_sk,
    cs.cs_item_sk AS item_sk,
    cs.cs_bill_addr_sk AS addr_sk,
    cs.cs_net_paid AS sales_amt,
    cr_ret.return_amt AS return_amt
  FROM catalog_sales cs
  LEFT JOIN (
    SELECT cr_item_sk, cr_order_number, SUM(cr_return_amount) AS return_amt
    FROM catalog_returns
    GROUP BY cr_item_sk, cr_order_number
  ) cr_ret
    ON cr_ret.cr_item_sk = cs.cs_item_sk
   AND cr_ret.cr_order_number = cs.cs_order_number
) s
JOIN item i ON i.i_item_sk = s.item_sk
JOIN customer_address ca ON ca.ca_address_sk = s.addr_sk
JOIN date_dim d ON d.d_date_sk = s.sold_date_sk
WHERE d.d_year = 2001
  AND (
    SELECT COUNT(*)
    FROM warehouse w
    WHERE w.w_state = ca.ca_state
  ) >= 2
GROUP BY
  d.d_year,
  s.channel,
  ca.ca_state,
  i.i_category
HAVING
  SUM(s.sales_amt) > CAST(100000 AS decimal(12,2))
  AND COALESCE(SUM(s.return_amt), 0) < SUM(s.sales_amt) * 0.1
ORDER BY
  d.d_year,
  s.channel,
  ca.ca_state,
  i.i_category
\end{tpcdsquery}

\begin{tpcdsquery}{A Query with Complexity 8}
WITH d AS (
  SELECT d_date_sk, d_year, d_qoy
  FROM date_dim
),
sales_channel AS (
  SELECT
    'store'::text AS channel,
    ss.ss_item_sk AS item_sk,
    d.d_year,
    d.d_qoy,
    SUM(ss.ss_quantity)::numeric AS qty_sold,
    SUM(ss.ss_net_paid) AS net_paid,
    SUM(
      COALESCE(
        CASE
          WHEN ib.ib_lower_bound IS NOT NULL AND ib.ib_upper_bound IS NOT NULL THEN ((ib.ib_lower_bound + ib.ib_upper_bound) / 2.0)::numeric
          WHEN ib.ib_lower_bound IS NOT NULL THEN ib.ib_lower_bound::numeric
          WHEN ib.ib_upper_bound IS NOT NULL THEN ib.ib_upper_bound::numeric
          ELSE NULL
        END,
        0::numeric
      ) * ss.ss_net_paid
    ) AS income_weighted_paid
  FROM store_sales ss
  JOIN d ON ss.ss_sold_date_sk = d.d_date_sk
  LEFT JOIN household_demographics hd ON ss.ss_hdemo_sk = hd.hd_demo_sk
  LEFT JOIN income_band ib ON hd.hd_income_band_sk = ib.ib_income_band_sk
  GROUP BY 1,2,3,4
  UNION ALL
  SELECT
    'web'::text AS channel,
    ws.ws_item_sk AS item_sk,
    d.d_year,
    d.d_qoy,
    SUM(ws.ws_quantity)::numeric AS qty_sold,
    SUM(ws.ws_net_paid) AS net_paid,
    SUM(
      COALESCE(
        CASE
          WHEN ib.ib_lower_bound IS NOT NULL AND ib.ib_upper_bound IS NOT NULL THEN ((ib.ib_lower_bound + ib.ib_upper_bound) / 2.0)::numeric
          WHEN ib.ib_lower_bound IS NOT NULL THEN ib.ib_lower_bound::numeric
          WHEN ib.ib_upper_bound IS NOT NULL THEN ib.ib_upper_bound::numeric
          ELSE NULL
        END,
        0::numeric
      ) * ws.ws_net_paid
    ) AS income_weighted_paid
  FROM web_sales ws
  JOIN d ON ws.ws_sold_date_sk = d.d_date_sk
  LEFT JOIN household_demographics hd ON ws.ws_bill_hdemo_sk = hd.hd_demo_sk
  LEFT JOIN income_band ib ON hd.hd_income_band_sk = ib.ib_income_band_sk
  GROUP BY 1,2,3,4
  UNION ALL
  SELECT
    'catalog'::text AS channel,
    cs.cs_item_sk AS item_sk,
    d.d_year,
    d.d_qoy,
    SUM(cs.cs_quantity)::numeric AS qty_sold,
    SUM(cs.cs_net_paid) AS net_paid,
    SUM(
      COALESCE(
        CASE
          WHEN ib.ib_lower_bound IS NOT NULL AND ib.ib_upper_bound IS NOT NULL THEN ((ib.ib_lower_bound + ib.ib_upper_bound) / 2.0)::numeric
          WHEN ib.ib_lower_bound IS NOT NULL THEN ib.ib_lower_bound::numeric
          WHEN ib.ib_upper_bound IS NOT NULL THEN ib.ib_upper_bound::numeric
          ELSE NULL
        END,
        0::numeric
      ) * cs.cs_net_paid
    ) AS income_weighted_paid
  FROM catalog_sales cs
  JOIN d ON cs.cs_sold_date_sk = d.d_date_sk
  LEFT JOIN household_demographics hd ON cs.cs_bill_hdemo_sk = hd.hd_demo_sk
  LEFT JOIN income_band ib ON hd.hd_income_band_sk = ib.ib_income_band_sk
  GROUP BY 1,2,3,4
),
returns_channel AS (
  SELECT
    'store'::text AS channel,
    sr.sr_item_sk AS item_sk,
    d.d_year,
    d.d_qoy,
    SUM(sr.sr_return_quantity)::numeric AS qty_ret,
    SUM(sr.sr_net_loss) AS net_loss
  FROM store_returns sr
  JOIN d ON sr.sr_returned_date_sk = d.d_date_sk
  GROUP BY 1,2,3,4
  UNION ALL
  SELECT
    'web'::text AS channel,
    wr.wr_item_sk AS item_sk,
    d.d_year,
    d.d_qoy,
    SUM(wr.wr_return_quantity)::numeric AS qty_ret,
    SUM(wr.wr_net_loss) AS net_loss
  FROM web_returns wr
  JOIN d ON wr.wr_returned_date_sk = d.d_date_sk
  GROUP BY 1,2,3,4
  UNION ALL
  SELECT
    'catalog'::text AS channel,
    cr.cr_item_sk AS item_sk,
    d.d_year,
    d.d_qoy,
    SUM(cr.cr_return_quantity)::numeric AS qty_ret,
    SUM(cr.cr_net_loss) AS net_loss
  FROM catalog_returns cr
  JOIN d ON cr.cr_returned_date_sk = d.d_date_sk
  GROUP BY 1,2,3,4
),
item_dim AS (
  SELECT i_item_sk AS item_sk, i_category, i_brand, i_manufact_id
  FROM item
),
combined AS (
  SELECT
    s.channel,
    i.i_category,
    i.i_brand,
    s.d_year,
    s.d_qoy,
    s.item_sk,
    s.qty_sold,
    s.net_paid,
    COALESCE(r.qty_ret, 0::numeric) AS qty_ret,
    COALESCE(r.net_loss, 0::numeric) AS net_loss,
    s.income_weighted_paid
  FROM sales_channel s
  JOIN item_dim i ON s.item_sk = i.item_sk
  LEFT JOIN returns_channel r
    ON r.channel = s.channel
   AND r.item_sk = s.item_sk
   AND r.d_year = s.d_year
   AND r.d_qoy = s.d_qoy
),
channel_cat_qtr AS (
  SELECT
    channel,
    i_category,
    d_year,
    d_qoy,
    SUM(qty_sold) AS qty_sold,
    SUM(net_paid) AS net_paid,
    SUM(qty_ret) AS qty_ret,
    SUM(net_loss) AS net_loss,
    SUM(income_weighted_paid) AS income_weighted_paid
  FROM combined
  GROUP BY 1,2,3,4
),
cat_qtr_totals AS (
  SELECT
    i_category,
    d_year,
    d_qoy,
    SUM(net_paid) AS total_net_paid,
    SUM(qty_sold) AS total_qty_sold,
    SUM(net_loss) AS total_net_loss,
    SUM(qty_ret) AS total_qty_ret,
    SUM(income_weighted_paid) AS total_income_weighted_paid
  FROM channel_cat_qtr
  GROUP BY 1,2,3
),
channel_metrics AS (
  SELECT
    ccq.channel,
    ccq.i_category,
    ccq.d_year,
    ccq.d_qoy,
    ccq.qty_sold,
    ccq.net_paid,
    ccq.qty_ret,
    ccq.net_loss,
    ccq.income_weighted_paid,
    CASE WHEN ccq.net_paid = 0 THEN NULL ELSE ccq.net_loss / ccq.net_paid END AS channel_loss_ratio,
    CASE WHEN ccq.qty_sold = 0 THEN NULL ELSE ccq.qty_ret / ccq.qty_sold END AS channel_return_rate,
    SUM(ccq.net_paid) OVER (PARTITION BY ccq.i_category, ccq.d_year, ccq.d_qoy) AS cat_qtr_net_paid,
    SUM(ccq.net_loss) OVER (PARTITION BY ccq.i_category, ccq.d_year, ccq.d_qoy) AS cat_qtr_net_loss
  FROM channel_cat_qtr ccq
),
channel_ranked AS (
  SELECT
    cm.*,
    CASE
      WHEN cm.cat_qtr_net_paid = 0 THEN NULL
      ELSE cm.net_paid / cm.cat_qtr_net_paid
    END AS channel_share,
    DENSE_RANK() OVER (PARTITION BY cm.i_category, cm.d_year, cm.d_qoy ORDER BY cm.channel_loss_ratio DESC NULLS LAST) AS loss_rank
  FROM channel_metrics cm
),
category_metrics AS (
  SELECT
    cqt.i_category,
    cqt.d_year,
    cqt.d_qoy,
    cqt.total_net_paid,
    cqt.total_qty_sold,
    cqt.total_net_loss,
    cqt.total_qty_ret,
    CASE WHEN cqt.total_net_paid = 0 THEN NULL ELSE cqt.total_income_weighted_paid / cqt.total_net_paid END AS weighted_avg_income,
    CASE WHEN cqt.total_net_paid = 0 THEN NULL ELSE cqt.total_net_loss / cqt.total_net_paid END AS category_loss_ratio,
    CASE WHEN cqt.total_qty_sold = 0 THEN NULL ELSE cqt.total_qty_ret / cqt.total_qty_sold END AS category_return_rate,
    AVG(CASE WHEN cqt.total_net_paid = 0 THEN NULL ELSE cqt.total_net_loss / cqt.total_net_paid END)
      OVER (PARTITION BY cqt.i_category ORDER BY cqt.d_year, cqt.d_qoy ROWS BETWEEN 3 PRECEDING AND CURRENT ROW) AS loss_ratio_ma4
  FROM cat_qtr_totals cqt
),
global_income AS (
  SELECT
    CASE WHEN SUM(net_paid) = 0 THEN NULL ELSE SUM(income_weighted_paid) / SUM(net_paid) END AS global_weighted_avg_income
  FROM channel_cat_qtr
)
SELECT
  cr.i_category,
  cr.d_year,
  cr.d_qoy,
  cr.channel,
  cr.channel_share,
  cr.channel_loss_ratio,
  cr.channel_return_rate,
  cr.loss_rank,
  cm.category_loss_ratio,
  cm.category_return_rate,
  cm.loss_ratio_ma4,
  cm.weighted_avg_income,
  gi.global_weighted_avg_income,
  CASE
    WHEN cm.weighted_avg_income IS NOT NULL AND gi.global_weighted_avg_income IS NOT NULL AND cm.weighted_avg_income > gi.global_weighted_avg_income THEN 'above_global'
    WHEN cm.weighted_avg_income IS NOT NULL AND gi.global_weighted_avg_income IS NOT NULL AND cm.weighted_avg_income < gi.global_weighted_avg_income THEN 'below_global'
    ELSE 'unknown'
  END AS income_position
FROM channel_ranked cr
JOIN category_metrics cm
  ON cm.i_category = cr.i_category
 AND cm.d_year = cr.d_year
 AND cm.d_qoy = cr.d_qoy
CROSS JOIN global_income gi
WHERE cr.loss_rank <= 3;
\end{tpcdsquery}

\clearpage
\onecolumn
\subsection{Prompts}
\subsubsection{Query Generation Prompts}
\label{sec:query_gen_prompts}
\begin{promptblock}{System Prompt}
You are a professor of database systems and SQL and your task is to generate complex SQL queries to probe the understanding of your students. You will be given a set of tables with their schemas, and you will generate a SQL query that matches the specified complexity level on a scale of 1-10, where 1 is simple queries with basic SELECT statements and 10 is highly complex queries with multiple nested subqueries, advanced window functions, CTEs, and intricate JOINs. Do not include any LIMIT clauses in your queries.

CRITICAL: All queries MUST be PostgreSQL compatible. Use only PostgreSQL syntax and functions. Lower complexity queries (1-3) should focus on basic operations (1 is only a select with a join and a simple where clause, while 3 is a select with a where clause and a 2/3 joins), medium complexity (4-7) should include JOINs and aggregations, and high complexity (8-10) should feature advanced techniques like window functions, complex subqueries, and sophisticated analytical operations.

Do not generate any text that is not SQL - do not add comments, explanations, or any other text.

You will return the SQL query within <query> and </query> tags.

\end{promptblock}
\vspace{0.6em}
\begin{promptblock}{User Prompt}
Please generate a SQL query with complexity level {complexity} (on a scale of 1-10) over the following tables: {tables}.
Tables have the following schema:
{schema}

IMPORTANT: The query MUST be PostgreSQL compatible. Use only PostgreSQL syntax and functions.

Match the specified complexity level:
- Complexity 1-2: Simple queries with basic SELECT statements, a join and simple WHERE clauses
- Complexity 3-4: SELECT with WHERE clause and 2-3 JOINs, start adding aggregations
- Complexity 5-6: even more JOINs between tables, GROUP BY, HAVING clauses, basic subqueries
- Complexity 7-8: Complex JOINs, window functions, CTEs, nested subqueries
- Complexity 9-10: Highly complex queries with multiple nested subqueries, advanced window functions, CTEs, and intricate JOINs

Do not include any LIMIT clauses in your queries.
Do not generate any text that is not SQL - do not add comments, explanations, or any other text.
Do not generate SQL queries that are only a simple SELECT. Even if the complexity is 1, you should have a join and a where clause.

CRITICAL CONSTRAINTS:
- Do NOT use WITHIN GROUP syntax as it is not supported yet
- Ensure all referenced fields actually exist in the tables being queried (avoid semantic errors)
- When using aggregations (avg, sum, etc.), ensure the aggregator can be applied to the target data type (e.g., avg cannot be applied to string types)
- Correlated scalar subqueries can only be used in Projection, Filter, Aggregate plan nodes
- Make sure you use PostgreSQL syntax and functions available in datafusion.

Remember to use the <query> and </query> tags to return the SQL query.

<query>

```sql

[some sql]

```
</query>

It is important you follow the format, otherwise the query will not be parsed correctly.
\end{promptblock}
\subsubsection{Plan Optimization Prompts}
\label{sec:plan_optim_prompts}
\label{prompt:opt_system}
\begin{promptblock}{System prompt}
You are an expert in Apache DataFusion physical execution plan optimization.
## Task
Analyze the CURRENT execution plan and generate JSON patches (RFC 6902) to optimize the execution plan structure, reducing execution cost while preserving query semantics. You should assume that the current plan can almost always be improved and must not be lazy: actively search for semantics-preserving structural changes instead of defaulting to making no changes.

## Input
- `structure`: Current execution plan to analyze and improve. This structure is correct.
- `succinct_table_info`: Table information for understanding the plan
- `query`: SQL query this plan executes

## Structure Format

The execution plan is a JSON object where:
- Keys are node IDs (e.g., "0", "1", "2")
- Values are node definitions with operation-specific fields
- Nodes reference other nodes via:
  - `"input": <node_id>` for single-input operations (filters, aggregates, etc.)
  - `"left": <node_id>` and `"right": <node_id>` for binary operations (joins)


## Succinct Table Info Format

The succinct table info is a JSON object that contains all the information about the tables in the plan.
Example structure fragment:
```json
{{
  "6": {{"hashJoin": {{"left": 7, "right": 26, "on": [...], "projection": [...]}}}},
  "7": {{"coalesceBatches": {{"input": 8, "targetBatchSize": 8192}}}},
  "26": {{"coalesceBatches": {{"input": 27, "targetBatchSize": 8192}}}}
}}

```

### Optimization steps

**Step 1:** cardinality estimation using semantic reasoning and domain knowledge.

Use your understanding of data semantics and real-world patterns to estimate cardinalities. Leverage semantic analysis of column names, table contexts, and filter predicates:

- **ParquetScan**: Use the provided row count from table statistics.
- **Filter nodes**: Analyze column semantics (temporal, ID, status, demographic columns) and filter predicates to estimate selectivity based on real-world data distributions. Consider table context (user behavior, business cycles, popularity patterns) when making estimates.
- **Join operations**: Estimate based on relationship semantics (foreign keys, many-to-many, temporal joins) and typical join ratios.

CRITICAL: Do not use defaultFilterSelectivity or other default values. Instead, perform semantic analysis of column names, table contexts, and filter predicates to make intelligent cardinality estimates based on real-world knowledge.

You should output your cardinality estimation of each node from bottom to top of the plan.

**Step 2**: use cardinality estimation to improve query plan

**1. Join-Side Selection**: Use your cardinality estimation to check the left and right input of the a hash join, and swap the input order if current left is larger than right, i.e., `left` should always be the smaller input.

Example: `"left": 7, "right": 26` → `"left": 26, "right": 7` if node 26 produces fewer rows than node 7

**2. Join Reordering**: In a multi-join query, join with lower cardinality should be performed first. For example, if relation A, B, C are joined on the same key, and we should first join A and C if A join C's cardinality is smaller than A join B. More formally, `(A ⨝ B) ⨝ C` → `(A ⨝ C) ⨝ B` if `|A ⨝ C| < |A ⨝ B|`

## Optimization invariants
1. Preserve all nodes, do not remove nodes or leave nodes that are not connected to something. There are no redundant nodes in the plan.
2. Maintain valid DAG topology (no cycles, valid references)
3. Update metadata when making structural changes - when swapping nodes or reordering joins, ensure that any associated metadata is also updated to reflect the new structure, including join keys/conditions and projection indexes (see below).

**Update Projection Index After Swapping Join Inputs**

When you swap the left and right inputs of a HashJoin, you MUST update the projection indexes to reflect the new schema order. The projection calculation follows this rule:
- If projection references a left field: use the index as-is
- If projection references a right field: offset by len(left_schema), i.e., `len(left_schema) + right_projection[i]`

**Example:**
```
Original:

- Left table: A [name, id] (schema length = 2)
- Right table: B [dept_name, budget] (schema length = 2)
- Join output: [name, budget]
- Projection: [0, 3] (name=0 from A, budget=1+2=3 from B)

After swapping left/right inputs:
- New left: B [dept_name, budget] (schema length = 2)
- New right: A [name, id] (schema length = 2)
- Join output: [name, budget]
- New projection: [2, 1] (name=0+2=2 from A, budget=1 from B)
```
To determine the schema length, you should recursively check the projection count from the input nodes.

** End Example **

## JSON Patch Operations
Use standard JSON Patch operations targeting the correct field names. **IMPORTANT**: When replacing objects, include ALL necessary fields including `index` fields - they may need to be updated/swapped based on the changes:
- `{{"op": "replace", "path": "/6/hashJoin/on/0/left", "value": {{"column": {{"name": "new_col"}}}}}}` - Update join condition after swap
- `{{"op": "replace", "path": "/6/hashJoin/on/0/right", "value": {{"column": {{"name": "other_col", "index": <correct_value>}}}}}}` - Update join condition after swap
- `{{"op": "replace", "path": "/4/aggregate/input", "value": 10}}` - Replace input of aggregate
- `{{"op": "move", "from": "/1", "path": "/0"}}` - Move entire node definition

By default, the JSON patch array you output should contain at least one operation that changes the plan structure. Returning an empty array `[]` (for example `<patch>[]</patch>`) is acceptable only in critical cases where, after careful and exhaustive analysis, you determine that no semantics-preserving structural change can reduce the execution cost.

Respond with a JSON patches array in the following format:

<patch>
[
  {{"op": "replace", "path": "/6/hashJoin/left", "value": .....}},
  {{"op": "replace", "path": "/6/hashJoin/right", "value": .....}}
]
</patch>
\end{promptblock}
\vspace{0.6em}
\label{prompt:opt_user}
\begin{promptblock}{User prompt}
<succinct_table_info>
{json.dumps(succinct_table_info, indent=2)}
</succinct_table_info>

<query>
{query}
</query>

<structure>
{json.dumps(structure, indent=2)}
</structure>
\end{promptblock}

\twocolumn
\clearpage

\end{document}